\documentclass[apjl]{emulateapj}
\usepackage{epsfig}

\def\beq{\begin{equation}}
\def\eeq{\end{equation}}

\begin{document}

\shorttitle {Solar Coronal Heating}
\shortauthors{D.~A.~Uzdensky}

\title{Fast Collisionless Reconnection Condition and 
Self-Organization of Solar Coronal Heating}
\author{Dmitri A.\ Uzdensky\altaffilmark{1}}
\altaffiltext{1}{Department of Astrophysical Sciences, Princeton University, 
Peyton Hall, Princeton, NJ 08544 --- Center for Magnetic Self-Organization
(CMSO); {\tt uzdensky@astro.princeton.edu}.}

\date{August 27, 2007}

\begin{abstract}
I propose that solar coronal heating is a self-regulating process 
that keeps the coronal plasma roughly marginally collisionless.
The self-regulating mechanism is based on the interplay of two effects.
First, plasma density controls coronal energy release via the transition 
between the slow collisional Sweet--Parker regime and the fast collisionless 
reconnection regime. This transition takes place when the Sweet--Parker
layer becomes thinner than the characteristic collisionless reconnection 
scale. I present a simple criterion for this transition in terms of the 
upstream plasma density ($n_e$), the reconnecting ($B_0$) and guide ($B_z$) 
magnetic field components, and the global length ($L$) of the reconnection 
layer: 
$L \lesssim 6\cdot 10^{9}\,{\rm cm}\ (n_e/10^{10}{\rm cm}^{-3})^{-3}\, 
(B_0/30\,{\rm G})^4\, (B_0/B_z)^2$. 
Next, coronal energy release by reconnection raises the ambient 
plasma density via chromospheric evaporation and this, in turn, temporarily 
inhibits subsequent reconnection involving the newly-reconnected loops.
Over time, however, radiative cooling gradually lowers the density again 
below the critical value and fast reconnection again becomes possible.
As a result, the density is highly inhomogeneous and intermittent but, 
statistically, does not deviate strongly from the critical value which
is comparable with the observed coronal density. Thus, in the long run,
the coronal heating process can be represented by repeating cycles that
consist of fast reconnection events (i.e., nanoflares), followed by rapid 
evaporation episodes, followed by relatively long periods (~$\lesssim$~1~hr) 
during which magnetic stresses build up and simultaneously the plasma cools 
down and precipitates.
\vspace{0.1 in}
\end{abstract}

\keywords{magnetic fields --- MHD --- Sun: corona --- Sun: magnetic fields
--- Sun: flares}


\section{Introduction}
\label{sec-intro}

In this paper I address some aspects of solar coronal heating 
(see Aschwanden~et~al.~2001b and Klimchuk~2006 for recent reviews) 
in the context of Parker's nanoflare model (Parker~1972, 1983, 1988). 
A more concise version of this work is presented in Uzdensky (2006, 2007).

Since the main heating process in the nanoflare model is magnetic 
reconnection, I first discuss what we have learned about reconnection 
in the past 20~years or so (see~\S~\ref{sec-reconnection}). I argue that, 
even though we still do not have a complete picture of reconnection, there 
now appears to be some consensus in the magnetic reconnection community
about some of its fundamental aspects. One of the main goals 
of this paper is to use this emerging knowledge to shed some new 
light on the old coronal heating problem. In this paper I purposefully 
adopt a rather conservative approach: I invoke only those very few results 
that seem to be relatively firmly established and try not to rely on those
details of reconnection physics that are still under vigorous debate. 
Specifically, there is now strong evidence coming from numerous 
numerical simulations and some laborious laboratory experiments
that there are two main modes of reconnection: slow Sweet--Parker
reconnection taking place in collisional plasmas where classical 
resistive MHD applies; and fast Petschek-like reconnection in
collisionless plasmas (in \S~\ref{subsec-2regimes}). The transition 
between these two regimes seems to be rather sharp. An approximate 
condition for this transition can be formulated as a relationship 
between the global length~$L$ of the reconnecting system and the 
electron collisional mean-free path inside the layer, 
$\lambda_{e,\rm mfp}$ (see \S~\ref{subsec-condition}). Furthermore, 
this condition can be cast in terms of the plasma density~$n$, 
the reconnecting magnetic field~$B_0$, and~$L$; namely, one can 
define a critical density $n_c(L,B_0)$ below which reconnection 
switches to the fast regime. In the strong-guide field case, $B_z\gg B_0$, 
the condition is modified; namely, the critical density for transition to 
fast reconnection is suppressed by a factor of order $(B_z\gg B_0)^{2/3}$ 
(see \S~\ref{subsec-guide-field}). In \S~\ref{subsec-caveats} I address 
various alternative ideas and caveats that may complicate the above 
simple picture.

I then apply these results to the active solar corona 
(in \S~\ref{sec-corona}), viewed withing the nanoflare
model. My main point here is that the corona should be 
regarded as a self-regulating machine keeping itself 
(in a statistical sense) around marginal collisionality. 
This conclusion comes from the interplay between the way
the plasma density controls reconnection via the above 
collisionless reconnection transition, and the way the 
coronal magnetic energy release due to reconnection in 
turn controls the ambient gas density via chromospheric
evaporation. The coronal heating process is then highly
intermittent and inhomogeneous; it can be thought of a 
sequence of characteristic energy-circulation cycles that
occur simultaneously in a broad range of spatial, temporal, 
and energy scales. Each such elementary cycle consists of 
several phases: (1) a fast reconnection event (a nanoflare)
causing (2) an evaporation episode filling the loop with hot
dense plasma, followed by (3) a longer period during which 
the magnetic stresses build up and the plasma density goes 
down due to slow radiative cooling (and thermal conduction). 
It is the collisionless reconnection condition that makes 
this scenario (in particular, the last phase) possible. 
Indeed, an important point here is that, even if a current 
sheet is formed, it will stay in the slow Sweet--Parker state 
if the density is large enough. This will continue for some 
time, until the reconnecting magnetic field becomes strong 
enough and/or the ambient plasma density becomes low enough 
for the system to transition to the fast collisionless 
reconnection regime. This enables a non-trivial amount
of free magnetic energy to be accumulated before a sudden
release. This energy amount, along with the characteristic 
time-scale between nanoflares and the characteristic coronal
density, is determined, statistically, by the balance between
the rate at which current sheets are created and amplified by
the photospheric footpoint motions and the efficiency of radiative 
cooling. The collisionless reconnection condition plays a key role 
in this process as a mediator and ultimately governs the statistical 
distribution of the coronal reconnection events (flares).

In \S~\ref{sec-discussion}, I discuss some of the open questions that, 
in my view, need to (and can) be addressed in the near future, in order
to see whether the physical picture put forward in this paper is correct
and what modifications should be made to improve it. This section is mostly 
targeted towards researchers doing numerical simulations of reconnection 
and also to experimentalists and observers. 
Finally, I present my conclusions in~\S~\ref{sec-conclusions}.

I also would like to make a clarifying remark about the terms 
``collisional'' and ``collisionless'' that will be used many 
times throughout this paper. Often, by ``collisionless'' one 
means a regime in which $\Omega_e > \nu_{e,i}$, i.e., when the 
electron gyro-frequency is higher than the electron-ion collision 
rate. Equivalently, this means that $\lambda_{e,\rm mfp}>\rho_e$. 
In this paper, on the other hand, the terms~``collisional'' and 
``collisionless'' will involve comparing~$\lambda_{e,\rm mfp}$ 
not with the gyro-radius, but instead with the system size~$L$, 
or, more precisely, with~$L$ divided by the square root of the 
ion/electron mass ratio. As will be explained in~\S\S~\ref
{subsec-condition}--\ref{subsec-T_e-n_e-SP} [specifically, 
see eq.~(\ref{eq-condition-2}) and the discussion at the end 
of \S~\ref{subsec-T_e-n_e-SP}], this choice of terminology turns 
out to be more relevant to the reconnection problem.


\section{Fast Collisionless Reconnection Condition}
\label{sec-reconnection}

\subsection{The Existence of Two Reconnection Regimes}
\label{subsec-2regimes}

Magnetic reconnection research started 50 years ago with the Sweet--Parker 
(SP) theory (Sweet~1958; Parker~1957) for solar flares. As was realized 
immediately at that time, this relatively simple and elegant theory could not 
reproduce the very short ($\sim 10^3$~sec) reconnection timescales required 
by flare observations; instead, it predicted reconnection times of order a 
few weeks or months. After several years, however, Petschek (1964) proposed 
a modification to the classic Sweet--Parker theory that apparently resulted 
in a much higher reconnection rate, thus eliminating the main contradiction 
with observations. He realized that the main bottleneck in the Sweet--Parker 
reconnection model is the need to have a reconnection layer that is both thin 
enough for the resistivity to be important and at the same time thick enough 
for the plasma to be able to flow out. The result of this compromise is the 
famous Sweet--Parker scaling for the thickness~$\delta_{\rm SP}$ of the 
reconnection layer and for the reconnection velocity~$v_{\rm rec}$:
\beq
{{\delta_{\rm SP}}\over L} = {{v_{\rm rec}}\over{V_A}} = 
\sqrt{\eta\over{L V_A}} \equiv S^{-1/2} \, ,
\label{eq-delta_SP}
\eeq
where $L$ is the global length of the reconnection layer, of order 
the system size, $\eta$ is the magnetic diffusivity, and $V_A$ is 
the Alfv\'en speed corresponding to the reconnecting (in-plane) 
magnetic field component. Since the Lundquist number in the solar 
corona is usually very large ($10^{12}$ or greater), the layer is 
very thin and the resulting reconnection rate is very small.
Furthermore, Petschek (1964) proposed that this difficulty can be 
circumvented if the reconnection region has a certain special structure: 
the famous Petschek configuration, with four slow-mode shocks attached to 
a small Sweet--Parker central diffusion region. As he showed, this structure 
gave an additional geometric factor that could lead to a much faster 
reconnection rate.

I would like to emphasize the special importance of Petschek's idea 
for reconnection in astrophysical systems, including the solar corona.
As I will discuss below, there are several local physical effects 
(e.g., Hall effect or anomalous resistivity) that can indeed broaden 
the reconnection layer and prevent it from collapsing down to the very 
small Sweet--Parker thickness~$\delta_{\rm SP}$. However, each of these 
processes comes with its own microscopic physical scale, e.g., the ion 
gyro-radius~$\rho_i$  or the ion collisionless skin-depth~$d_i \equiv 
c/\omega_{pi}$. These scales are determined by the {\it local} values 
of the basic  plasma parameters inside the reconnection layer, such as 
magnetic field, density, or temperature. The important point is that 
they know nothing about the overall {\it global} system size~$L$. 
{\it Astronomical systems are often astronomically large}, however. 
That is, $L$ is typically much larger than any microscopic physical 
scale~$\delta$ and than the Sweet--Parker layer thickness (which is 
a hybrid length-scale).
Therefore, any simple Sweet--Parker-like analysis would give a reconnection 
rate~$v_{\rm rec}/V_A$ scaling as~$\delta/L \ll 1$; this would not be rapid 
enough to be of any practical interest. Thus, we come to {\bf Conclusion~I}: 
{\it irrespective of the actual microphysics 
inside the layer, Petschek's mechanism, or a variation thereof, is absolutely 
necessary for sufficiently fast astrophysical reconnection}, including solar 
flares (and even micro- and nanoflares).


Because Petschek's model was able to reproduce the very short observed 
flare timescales, it quickly became very popular and people have believed 
in it for the next 20 years. With the advent of computer simulations, 
however, the original Petschek model came under fire. In particular, 
several numerical studies (e.g., Biskamp~1986; Scholer~1989; Ugai~1992, 1999;
Ma \& Bhattacharjee 1996; Uzdensky \& Kulsrud~1998, 2000; Erkaev~et~al. 2000, 
2001; Biskamp \& Schwarz 2001; Birn~et~al.~2001; Malyshkin et al.~2005; 
Cassak~2005, 2006) 
showed that in resistive MHD with uniform resistivity (and, by inference, 
with resistivity that is a smooth function of plasma parameters, e.g., 
Spitzer, see Biskamp \& Schwarz~2001) Petschek's mechanism fails and 
Sweet--Parker scaling applies instead. The basic physical reason for 
this, as has been elucidated analytically by Kulsrud (2001), (see also 
Uzdensky \& Kulsrud~2000 and Malyshkin~et~al. 2005) is that the transverse 
(i.e., perpendicular to the current sheet) magnetic field component, which 
is necessary to sustain Petschek's shocks, is swept away by the Alfv\'enic 
flow out of the layer so rapidly that it cannot be regenerated by the 
nonuniform resistive merging of the reconnecting magnetic field component.

In addition, a strong evidence for the existence of a slow Sweet-Parker 
reconnection mode in collisional resistive-MHD plasmas has been obtained
from recent laboratory studies in the Magnetic Reconnection Experiment 
(MRX; Yamada~et~al.\ 1997) in the high-collisionality regime (Ji~et~al.~1998; 
Trintchouk~et~al.~2003; Kuritsin~et~al. 2006).

This enables us to draw {\bf Conclusion~II}: 
{\it In the collisional regime, when classical resistive MHD applies, 
Petschek's fast reconnection mechanism does not work; slow Sweet--Parker
reconnection process takes place instead.}


A very natural question to ask next is whether fast Petschek-like 
reconnection possible in a collisionless plasma, where resistive 
MHD does not apply. There is now a growing consensus that the answer 
to this question is YES. In Solar and Space physics, of course, there 
has long been plentiful observational evidence for fast collisionless 
reconnection, in solar flares (e.g., Tsuneta~et~al.\ 1984, 1992, 1996; 
Masuda~et~al.\ 1994, 1995; Shibata~1995, 1996; Yokoyama~et~al. 2001), 
and in the Earth magnetosphere, both in the magnetopause (e.g., Mozer~et~al. 
2002, 2003, 2004) and in the magnetotail (Oieroset~et~al.\ 2001; 
Nagai~et~al. 2001). The evidence for fast collisionless reconnection 
has been further significantly strengthened by recent laboratory measurements 
in the MRX (Ji~et~al.\ 1998, 2004; Yamada~et~al.\ 2006) and other devices 
(e.g., Brown~1999; Egedal~et~al.~2007; Brown~et~al. 2006).
These laboratory measurements, however, have not been able to elucidate 
the special role of the Petschek mechanism in accelerating reconnection. 
On the other hand, over the past decade or so, several theoretical and 
numerical studies have indicated that fast reconnection enhanced by the 
Petschek mechanism (or a variation thereof) does indeed take place in 
the collisionless regime. Moreover, it appears that there may be two 
mechanisms of collisionless reconnection. Physically, these two mechanisms
are very different from each other; nevertheless, they both appear to lead 
to the establishment of a Petschek-like configuration, which enhances the 
reconnection rate. The two regimes in question are:

\begin{itemize}
\item{{\it Hall-MHD reconnection}, involving two-fluid effects 
in a {\it laminar} flow configuration (e.g., Mandt et al. 1994; 
Kleva et al. 1995; Biskamp et al. 1995, 1997; Ma \& Bhattacharjee 1996; 
Lottermoser \& Scholer 1997; Shay~et~al. 1998, 1999, 2001; Hesse et al. 1999; 
Bhattacharjee et al. 1999, 2001; Birn~et~al.~2001; Rogers et al. 2001; 
Pritchett 2001; Breslau \& Jardin 2003; Cassak et al. 2005, 2006; 
Daughton~et~al. 2006).
} 

\item{Spatially-localized {\it anomalous resistivity} due to plasma 
micro-turbulence that is excited when the current density exceeds a
certain threshold (e.g., Coppi \& Friedland 1971; Smith \& Priest 1972; 
Coroniti \& Eviatar 1977; Kulsrud 2001, 2005; Kulsrud~et~al. 2005). 
This seems to lead to a Petschek configuration with the length of 
the inner diffusion region being on the order of the resistivity 
localization scale 
(e.g., Ugai \& Tsuda 1977; Sato \& Hayashi 1979; Ugai~1986, 1992, 1999;
Scholer~1989; Biskamp \& Schwarz 2001; Erkaev~et~al. 2001; Kulsrud~2001,
2005; Uzdensky 2003; Malyshkin~et~al. 2005). 
The corresponding reconnection rate is then fast enough to explain 
the relatively-short timescales observed in solar flares (e.g., Kulsrud 2001, 
2005; Uzdensky~2003).
}
\end{itemize}

Recent experimental evidence from the MRX experiment indicates that 
both anomalous resistivity (e.g, Ji~et~al. 2004) and Hall effect 
(Ren~et~al.\ 2005; Yamada~et~al.\ 2006; Brown~et~al.\ 2006; 
Egedal~et~al.\ 2007) are present and may be important in collisionless 
reconnection. However, at present, it is still not clear which one of 
these two mechanisms works in a given physical situation (if at all). 
Also not known is whether these two mechanisms can operate simultaneously 
in a given system and how they interact with each other.

In any case, from the point of view of our present discussion,
the important thing is that one does get a Petschek-enhanced 
fast reconnection if the plasma is collisionless (in the sense 
that will be specified below, see \S~\ref{subsec-condition}). 
Thus, we can draw {\bf Conclusion~III}:
{\it a Petschek-enhanced fast reconnection does happen in 
the collisionless regime.}

To sum up, there are two regimes of astrophysical magnetic reconnection:
slow Sweet--Parker reconnection in resistive-MHD with classical 
collisional resistivity, and fast Petschek-like reconnection in 
collisionless plasmas.
In other words, in order for reconnection to be fast, it needs Petschek's 
mechanism to operate and that in turn requires the reconnection layer to 
be collisionless.

In fact, whenever we observe violent and rapid energetic phenomena that 
we interpret as reconnection, it is always in relatively tenuous plasmas. 
I am not aware of any counter-examples and would be would be very interested 
in learning about them. Is there any evidence for fast large-scale 
reconnection events in collisional astrophysical environments?


\subsection{The Fast Reconnection Condition: Zero Guide Field Case}
\label{subsec-condition}

How can one quantify the transition between the two regimes?
For clarity, let us first consider the case with zero guide field, $B_z=0$. 
[In the $B_z\neq 0$ case some of the arguments and results presented
in this section will be somewhat modified; however, the main concepts 
and conclusions will remain similar, as we show in the next section 
(see~\S~\ref{subsec-guide-field}).]

When there is no guide field (strictly anti-parallel field merging),
the condition for fast reconnection can be formulated roughly as 
(Kulsrud~2001, 2005; Uzdensky~2003; Cassak~et~al.~2005; Yamada~et~al.~2006)
\beq
\delta_{\rm SP} < d_i \equiv {c\over{\omega_{pi}}} \, .
\label{eq-condition-1}
\eeq
What this condition means is the following. As a reconnection layer forms, 
its thickness~$\delta$ becomes smaller and smaller. If condition~(\ref
{eq-condition-1}) is not satisfied, then this thinning saturates at 
$\delta=\delta_{\rm SP}$; then, reconnection proceeds in the slow 
Sweet--Parker regime. On the other hand, if condition~(\ref{eq-condition-1}) 
is satisfied, then various two-fluid and/or kinetic effects kick in 
as soon as~$\delta$ drops down to about~$d_i$ or so, i.e., well before 
the resistive term becomes important. Then, the reconnection processes 
necessarily involves collisionless, non-classical-resistive-MHD physics.
This results in a rapid increase in the reconnection rate, as has been
recently documented experimentally in the MRX (Yamada~et~al. 2006).

Note that, by construction, $\delta_{\rm SP}$ that enters in 
equation~(\ref{eq-condition-1}) is to be calculated using the 
classical Spitzer resistivity. Therefore, and since we are talking 
about collisional and collisionless reconnection, it is instructive 
to express the above condition in terms of the classical electron 
mean free path due to Coulomb collisions, $\lambda_{e,\rm mfp}$.
It is pretty straight-forward to show (Yamada~et~al. 2006) that
\beq
{{\delta_{\rm SP}}\over{d_i}} \sim 
\biggl({L\over{\lambda_{e,\rm mfp}}}\biggr)^{1/2}\, 
\biggl(\beta_e\,{m_e\over{m_i}}\biggr)^{1/4} \, ,
\label{eq-yamada-2006}
\eeq
Here, $\beta_e$ is the ratio of the electron pressure inside the layer 
to the pressure of the reconnecting magnetic field component ($B_0^2/8\pi$)
outside the layer. 
Thus, using equation~(\ref{eq-condition-1}), we see that 
a reconnection layer is collisionless when
\beq
L < L_c \equiv \lambda_{e,\rm mfp}\, \sqrt{m_i/{\beta_e m_e}} 
\simeq  40\, \beta_e^{-1/2}\, \lambda_{e,\rm mfp} \, .
\label{eq-condition-2}
\eeq
The collisionless reconnection condition in a similar form has been 
first published by Yamada~et~al. (2006). Their condition actually 
differs from equation~(\ref{eq-condition-2}) by a factor of~2. 
Since the discussion here is very qualitative, I regard this difference 
as unessential. Moreover, to make the present discussion more clear, I 
systematically ignore numerical factors of order~1 everywhere in this 
paper.

We can go a little bit further. The mean free path $\lambda_{e,\rm mfp}$ 
can be written as 
\beq
\lambda_{e,\rm mfp} \sim {{k_B^2 T_e^2}\over{n_e e^4 \log\Lambda}} \simeq
7\cdot 10^{7}{\rm cm}\, n_{10}^{-1}\, T_7^2 \, ,
\label{eq-lambda-1}
\eeq
where we set the Coulomb logarithm equal to~20 and where~$n_{10}$ and~$T_7$ 
are the central layer electron density~$n_e$ and temperature~$T_e$ given in 
units of $10^{10}\, {\rm cm}^{-3}$ and~$10^7$~K, respectively. 
Combining equations~(\ref{eq-condition-2}) and~(\ref{eq-lambda-1}), 
the criterion 
for fast collisionless reconnection can now be formulated as a condition 
on the layer's length~$L$ in terms of the values of~$n_e$ and~$T_e$ at 
the center of the Sweet--Parker reconnection layer:
\beq
L < L_c \equiv 40\, \beta_e^{-1/2}\, \lambda_{e,\rm mfp} 
\simeq 3\cdot 10^{9}{\rm cm}\ \beta_e^{-1/2}\, n_{10}^{-1}\, T_7^2 \, .
\label{eq-condition-3}
\eeq

Notice that the critical layer length~$L_c$ strongly depends 
on the central electron temperature~$T_e$. The main cause for 
this high sensitivity is the strong temperature dependence of 
the electron mean free path and hence Spitzer resistivity. 
It is therefore very important to figure out what the electron 
temperature inside the Sweet--Parker reconnection layer should 
be. In a situation where the ambient (upstream) pressure is 
already high, $\beta_{\rm upstream} \geq 1$, the conversion 
of magnetic energy to plasma via reconnection is not going 
to affect the plasma temperature and density considerably. 
In this case, both the plasma temperature and the density 
should be roughly uniform across the layer; then one can 
just substitute their upstream values into equation~(\ref
{eq-condition-3}). The story then ends here. 

However, in this paper we will be mostly interested in environments 
that are (globally) magnetically-dominated (and hence almost force-free), 
such as the solar corona and coronae of other stars and accretion disks. 
In this case, the ambient thermal pressure is smaller than the reconnecting 
magnetic field pressure. It is then important to distinguish the temperature 
inside the layer from the ambient coronal temperature (which is much lower). 
The simplest first step towards determining~$T_e$ is to note that, in the 
case with zero guide field, the condition of pressure balance across the 
layer dictates that
\beq
\beta \equiv {{8\pi n_e k_B (T_e+T_i)}\over{B_0^2}} = 1 \, .
\label{eq-pressure-balance}
\eeq
Next, assuming that~$T_e$ inside a Sweet--Parker reconnection layer is 
approximately equal to (and in any case not much smaller than) the ion 
temperature~$T_i$ --- the self-consistency of this assumption will be 
demonstrated at the end of this section --- we see that~$\beta_e$ should 
generally be close to~1/2 (and in any case of order~1). 

Then, using the pressure-balance condition~(\ref{eq-pressure-balance}), 
we can express the central electron temperature in terms of~$B_0$ and~$n_e$ 
as
\beq
T_e = {{B_0^2/8\pi}\over{2k_B n_e}} \simeq
1.4 \cdot 10^7\, {\rm K} \, B_{1.5}^2 \, n_{10}^{-1} \, ,
\label{eq-T_e}
\eeq
where $B_{1.5}$ is the outside magnetic field $B_0$ expressed in units 
of~30~G. Upon substituting this expression into equation~(\ref{eq-lambda-1}), 
we get
\beq
\lambda_{e,\rm mfp} \simeq 1.5\cdot 10^8{\rm cm}\, n_{10}^{-3}\, B_{1.5}^4 \, ,
\label{eq-lambda-2}
\eeq
and so the fast collisionless reconnection condition can be written 
in the final form as
\beq
L < L_c(n,B_0) \simeq 6\cdot 10^{9}{\rm cm}\, n_{10}^{-3}\, B_{1.5}^4 \,  .
\label{eq-L_c-final}
\eeq
We see that the condition $L< L_c$ is often satisfied in the solar corona
(e.g., in solar flares) and is definitely always satisfied in the Earth 
magnetosphere.


\subsection{Plasma Temperature and Density inside the Sweet--Parker 
Reconnection Layer}
\label{subsec-T_e-n_e-SP}

Note that the density that enters in the above expressions is, by 
definition, the density at the center of the reconnection layer.
In general, it is not known {\it a priori} and we would like, instead, 
to get expressions involving the background (upstream) density.
Equation~(\ref{eq-pressure-balance}) gives us one relationship 
between the two desired quantities (central~$n_e$ and~$T_e$).
However, we need to know both of them and by itself, equation~(\ref
{eq-pressure-balance}) does not tell us whether the central pressure
is increased (from a very low ambient value) to the required 
equilibrium level by raising the density or the temperature.
I will now argue, however, that, in the regime that is relevant 
here, the latter is more likely to be the case, i.e., that the 
pressure increase is mostly due to the increased temperature,
whereas the density should change relatively little.

Indeed, as is evident from the above discussion, the main
reason for the strong density dependence in equation~(\ref{eq-L_c-final})
is the $T_e$-dependence of the classical Spitzer resistivity, combined 
with the pressure balance condition~(\ref{eq-pressure-balance}). 
Therefore, what we are mostly interested in here are the central
electron density and temperature in the context of the resistive
Sweet--Parker theory. To determine them, we need more detailed 
information. Namely, we need to consider one more equation that 
we have not discussed up until now --- the equation of energy 
conservation. In general this equation should take into account
Ohmic heating and all the relevant loss mechanisms, such as advection, 
radiation, and electron thermal conduction. 

Let us consider a fluid element as it travels through the reconnection layer.
During its transit, the fluid element is subject to Ohmic heating and to all 
the above loss mechanisms. The characteristic time for the advection term is 
of course just the Alfv\'en transit time~$\tau_A\sim L/V_A$ --- the time a 
fluid element spends inside the layer. During this time the element is 
continuously heated by Ohmic heating, whose rate per unit volume can be 
estimated within the Sweet--Parker theory as 
\beq
\eta' j^2 \sim \eta' \, \biggl({cB_0\over{4\pi\delta}}\biggr)^2 \sim
{{B_0^2}\over{4\pi}} \, {\eta \over{\delta^2}}  = 
{{B_0^2}\over{4\pi}} \, {\eta S\over{L^2}} =
{{{B_0^2}\over{4\pi}}} \, {L\over{V_A}} \, ,
\eeq
where the magnetic diffusivity~$\eta$ is related to the resistivity~$\eta'$
via $\eta = \eta' c^2/4\pi$.%
\footnote
{I would like to add one extra word of caution
here. The Sweet--Parker theory that we are relying on here does not take into 
account the large {\it variation} of the Spitzer resistivity across the layer 
as the electron temperature increases by a large factor with respect to its 
relatively low ambient value. This brings in an extra uncertainty into our 
estimates.}
Thus, during its transit, the fluid element acquires just the right 
amount of heat to raise its temperature to about the equipartition 
level given by equation~(\ref{eq-T_e}) (Uzdensky~2003). We now need 
to see under what conditions one can ignore various loss mechanisms, 
namely, radiation and thermal conduction.

First, it is easy to see that, in the case of the solar corona, 
the radiative losses are indeed negligible on the time scales 
of interest ($\tau_A$). Indeed, the radiative cooling time is 
determined by the cooling function~$Q(T)$ 
via
\beq
\tau_{\rm rad} \sim {{2 c_v n k_B T}\over{n^2 Q(T)}} = 
{{3 k_B T}\over{n Q(T)}} \, .
\label{eq-tau_rad}
\eeq
In the temperature range of interest to the solar corona, 
$10^6\ {\rm K} \leq T_e \leq 10^8\ {\rm K}$, the cooling 
function varies between $10^{-23}$ and~$10^{-22}$~erg~sec$^{-1}$~cm$^3$
(e.g., Rosner~et~al.\ 1978; Priest~1984, p.~88; Cook~et~al. 1989). Then, 
taking $n = 10^{10}\ {\rm cm}^{-3}$, the typical radiative cooling times 
are of order 10~min (for~$T=2\cdot 10^6$~K) and longer (up to many days 
for~$T=10^8$~K). In any case, this is much longer than the characteristic 
Alfv\'en crossing time~$\tau_A$ which is usually a few seconds.

On the other hand, the smallness of the radiative losses in other 
astrophysical situations cannot be taken for granted, especially 
in High-Energy Astrophysics applications, such as coronae of accretion 
disks around black holes and neutron stars and also gamma-ray bursts. 
If radiative losses are important, then the above estimates for the 
central temperature would no longer apply and correspondingly the 
condition for transition to the fast reconnection regime would have 
to be modified.

Next, going back to the solar corona, we need to consider the effect 
of electron thermal conduction on the central temperature. Note that 
electrons are still well magnetized throughout most of the reconnection 
layer (this will be even more so in the presence of a guide field).
Therefore, the electron thermal conduction is highly anisotropic, 
and so we shall estimate the effects of parallel and perpendicular
thermal conduction separately. We shall start with the heat conduction 
along the magnetic field. Since we interested in the collisional case, 
$\lambda_{e,\rm mfp} \ll L$, the parallel heat transport is in 
the diffusive regime and can be described by a one-dimensional 
random walk of electrons along the magnetic field. The characteristic 
parallel thermal conduction time-scale can then be estimated as 
\beq
\tau_{e,\rm cond,\parallel} \sim 
{{\lambda_{e,\rm mfp}}\over{v_{e,\rm th}}} \, 
\biggl({L\over{\lambda_{e,\rm mfp}}}\biggr)^2
= {L\over{v_{e,\rm th}}} \, {L\over{\lambda_{e,\rm mfp}}}  
\sim \tau_A \  \sqrt{\biggl({m_e\over{\beta m_i}}\biggr)} \  
{L\over{\lambda_{e,\rm mfp}}} \, ,
\eeq
where $\tau_A$ and $\beta$ correspond to the reconnecting field~$B_0$.
Using expression~(\ref{eq-condition-2}), we can write this expression 
as
\beq
\tau_{e,\rm cond,\parallel} \sim \tau_A\ {L\over{L_c}}\, \beta^{-1} \, .
\eeq

In the magnetically-dominated environment of the solar corona,
where the thermal pressure outside of the reconnection region 
is small, we expect $\beta\sim 1$; it can conceivably be 
smaller, but we never expect it to be significantly higher than~1.
Therefore, we estimate that
\beq
\tau_{e,\rm cond,\parallel} \geq \tau_A \ {L\over{L_c}} \, .
\eeq

Thus, we see that if we are in the regime of interest --- that is,
in the expected regime of applicability of the collisional Sweet--Parker
theory, $L>L_c$, --- then $\tau_{e,\rm cond,\parallel}>\tau_A$ and 
hence the energy losses due to parallel electron thermal conduction 
are not important.

A similar line of arguments can be made to show that the characteristic
time for perpendicular (across the reconnection layer) electron thermal 
conduction is in fact automatically always comparable to the Alfv\'en
crossing time. Indeed, the cross-field electron conduction timescale in
the Sweet--Parker regime can be estimated as
\beq
\tau_{e,\rm cond,\perp} = {{\delta_{\rm SP}^2}\over{D_{e,\perp}}} \, ,
\eeq
where $D_{e,\perp}\sim \rho_e^2\, v_{e,\rm th}/\lambda_{e,\rm mfp}$ 
is the classical electron collisional diffusivity across the magnetic 
field. Using $\beta_e \sim 1$, we see that $\tau_{e,\rm cond,\perp}$ 
should scale as 
\beq
\tau_{e,\rm cond,\perp} \sim \tau_A\ {{\delta_{\rm SP}^2}\over{d_i^2}}\
\sqrt{m_i\over{m_e}}\ {{\lambda_{e,\rm mfp}}\over{L}} \, .
\eeq
Upon plugging equation~(\ref{eq-yamada-2006}) into this expression, 
we immediately find that all of the factors on the right-hand-side 
cancel and we simply see that 
\beq
\tau_{e,\rm cond,\perp} \sim \tau_A \, ,
\eeq
i.e., that the two timescales are automatically comparable to each other.

Combined with our previous result regarding the effect of
the parallel thermal conduction, we conclude that energy 
losses due to both parallel and perpendicular electron 
thermal conduction out of the layer are, at best, only 
marginally important in the collisional Sweet--Parker 
regime. Heat losses are still present but are not likely
to cause a decrease in the central electron temperature by 
more than a factor of order unity. This means that the jump
in the plasma pressure, required to maintain the pressure 
balance with the outside reconnecting magnetic field, should 
be attributed mostly to the increase in the plasma temperature
(due to Ohmic heating), whereas the density should not vary across
the layer by more than a factor of a few. Thus, for our rough
estimates, the density that enters in our estimate~(\ref{eq-T_e}) 
for the central temperature can be taken to be the ambient plasma density. 
As long as we are in the collisional regime, $L>L_c$, this result is 
consistent with the energy balance inside the layer that includes 
Ohmic heating, heat advection by the bulk plasma outflow, and electron 
thermal transport. As we shall see in the next section, this will be
true even in the presence of a guide field.

It is also interesting to check that the assumption~$T_i\simeq T_e$
is not strongly violated. In principle, this could be a worry, since 
the collisional electron-ion energy-equilibration rate is suppressed 
by the large mass ratio; hence, in general, the electron and ion 
temperatures in the layer need not be equal. For example, ions 
might have been much hotter than the electrons and hence might 
have provided the bulk of the pressure support against the outside 
magnetic field. The electron temperature in this case would have been 
far below the equipartition value (about~$10^7$~K). Correspondingly, 
the electron mean-free path would be much lower than that given by 
equation~(\ref{eq-lambda-2}).

The electron-ion temperature equilibration time, $\tau_{\rm EQ}$, is 
by a factor of $m_i/m_e$ longer than the electron collision time, 
$\tau_e$. Making use of the the pressure balance $c_s\sim (T_e/m_i)^{1/2} 
\sim V_A$, we can express $\tau_e$ in terms of $\lambda_{e,\rm mfp}$ and 
the Alfv\'en speed as 
$\tau_e \simeq \lambda_{e,\rm mfp}/v_{e, \rm th} \sim 
(\lambda_{e,\rm mfp}/c_s)\, \sqrt{m_e/m_i} \sim 
(\lambda_{e,\rm mfp}/V_A)\, \sqrt{m_e/m_i}$.
Then, we can compare $\tau_{\rm EQ}$ with the Alfv\'en crossing time, 
which is the characteristic time a a fluid element spends inside 
the layer:
\beq
\tau_{\rm EQ} \sim \tau_A\,  {{\lambda_{e,\rm mfp}}\over{L}}\,  
\sqrt{m_i\over{m_e}}  \sim \tau_A \, {L_c\over{L}} \, .
\eeq

Thus we see that, if we are in the collisional regime as defined by 
equation~(\ref{eq-condition-2}), the electrons and ions experience 
enough collisions with each other to equalize their temperatures 
while they transit through the layer. Thus, $T_i\simeq T_e$ should 
be a decent approximation in the Sweet--Parker regime.

Note that the plasma density also enters the collisionless reconnection 
condition via~$d_i$. The straight-forward logic of the above picture 
dictates that this quantity is to be estimated in the collisional regime.
However, it is plausible that a corresponding estimate based on the 
collisionless regime also has some relevance; in this case it is interesting 
to compare the two.  
The task of estimating the central~$n_e$ in the collisionless regime 
is more difficult than estimating it in the Sweet--Parker regime, since
there are many uncertainties here. However, it is also less important, 
since most of the strong density dependence in equation~(\ref{eq-L_c-final}) 
comes from the the steep temperature (and hence, indirectly, density) 
dependence of the Spitzer resistivity. 
Nevertheless, we still shall try to discuss the question of whether the 
density can vary significantly across a collisionless reconnection layer.
If the fast collisionless reconnection is due to anomalous resistivity,
then the effective~$\lambda_{e,\rm mfp}$ is determined by wave-particle 
collisions and hence will be relatively short. This will result, again, 
in suppression of parallel electron thermal conduction and hence in high 
temperature and correspondingly, according to the pressure balance 
condition~(\ref{eq-pressure-balance}), in a relatively mild density change
with respect to the upstream value (see also Uzdensky~2003). Then the above 
estimates are likely to be still valid (although in this case one might have 
to deal with a possibility that~$T_i \gg T_e$).

On the other hand, if the fast reconnection is due to the Hall effect, 
then we also may expect the density to be roughly uniform across the layer. 
This is because the ions become demagnetized on scales less than~$d_i$ 
and hence cannot develop structures on smaller scales.
The dynamics of ions inside the reconnection layer can be summarized
by noting that they are pulled directly into the layer by the bipolar 
electric field, which accelerates them up to the Alfv\'en speed, and 
then are ejected out along the layer by the Lorentz force associated 
with the quadrupole out-of-plane magnetic field (e.g., Uzdensky \& 
Kulsrud 2006). At any rate, one does not expect to see a significant 
variation of ion, and hence electron, density across the Hall reconnection 
layer (apart from the $d_i$-thick density depletion layers that run along 
the separatrices, as reported by Shay~et~al.~2001).

I also would like to make a little side remark about my use of terms here.
In this paper I use the terms ``collisional'' and ``collisionless'' a lot 
and I would like to clarify what I mean by them. Notice that the 
collisionless reconnection condition~(\ref{eq-condition-2}) is different 
from the customary usage of the term ``collisionless''. Often, by 
``collisionless'' one means a regime in which $\Omega_e > \nu_{e,i}$, 
i.e., when the electron gyro-frequency is faster than the electron-ion 
collision rate. Equivalently, this means that $\rho_e < \lambda_{e,\rm mfp}$. 
In this paper, on the other hand, the terms~``collisional'' and 
``collisionless'' involve comparing~$\lambda_{e,\rm mfp}$ not with 
the electron gyro-radius, but instead with the system size~$L$ 
divided by the square root of the ion/electron mass ratio, as 
seen from~(\ref{eq-condition-2}).

Finally, I would like to make a comment about the use of the term
``collisionless''. As I mentioned in~\S~\ref{sec-intro}, in the
present paper this term is understood in the sense of equation~(\ref
{eq-condition-2}). That is, a system is called ``collisionless'' when
$\lambda_{e,\rm mfp}\gtrsim L/40$, roughly speaking. This definition
is different from the condition $\lambda_{e,\rm mfp}>\rho_e$ that is 
often used. Note that, when $\lambda_{e,\rm mfp}>\rho_e$, or, equivalently,
$\Omega_e>\nu_{e,i}$, then the Hall term in the generalized Ohm law formally 
cannot be neglected. In plasmas of interest to this paper, e.g., in the solar 
corona, this condition is always very well satisfied. 
In this sense, the coronal plasma is definitely always collisionless. 
Our collisionless reconnection condition~(\ref{eq-condition-1}), on 
the other hand, is not always satisfied.
In fact, there is a large and astrophysically-important region in parameter
space where~$\delta_{\rm SP}>d_i$ but where the plasma is nevertheless
collisionless in the sense that $\Omega_e > \nu_{e,i}$, and where therefore
the Hall effect may be significant (see also Cassak~et~al. 2005). 
In the above discussion I have assumed that, in this intermediate regime, 
reconnection still proceeds slowly, in the purely resistive Sweet--Parker 
mode, as suggested by numerous resistive-MHD simulations. One should be 
aware, however, that most of these simulations completely ignore the Hall 
effect; thus, it is not clear to what extent their results can be relied 
upon in this intermediate regime. In fact, the recent numerical work by 
Cassak~et~al. (2005), who include both resistivity and Hall effect, 
indicates that both slow Sweet--Parker and fast Hall modes may operate 
in the intermediate regime, depending on the previous history of the 
system. In particular, they observe that, when one considers a gradual
initial thinning of the reconnection layer, the system always stays in the
slow regime, even if $\Omega_e > \nu_{e,i}$, as long as~$\delta_{\rm SP}>d_i$.
Since it is exactly this scenario that is of interest in the context of our
model, we feel we can use condition~(\ref{eq-condition-1}) with confidence.


\subsection{Reconnection with a Guide Field}
\label{subsec-guide-field}

In the real world the zero-guide-field case considered in the
previous section is rare. It may be encountered in some special
circumstances; for example it may be relevant to reconnection 
in the Earth magnetotail. Generally, however, there is some guide 
field, $B_z\neq 0$, present. This is especially so in the context 
of the nanoflare model for heating of solar coronal loops by braiding 
of individual elemental flux strands that comprise a larger loop. 
In this case the guide field (i.e., the axial field along the general 
loop direction) is always present 
and is generally much stronger than the reconnecting field~$B_0$. 
Another situation where guide field is dominant over the reconnecting 
magnetic field is laboratory plasma devices, such as tokamaks.
Note that reconnection releases the energy of the weaker $B_0$-field; 
the energy of the guide field is not available. However, the presence 
of a guide field, especially if it is strong, may, in principle, change 
the structure of the reconnection layer. Therefore, in this section we 
consider the effect of the guide field on our collisionless reconnection 
condition.

The presence of the guide field modifies the collisionless reconnection
condition. This is because, whereas the characteristic layer thickness 
in resistive MHD is not affected by the guide field and is still given 
by the same Sweet--Parker scaling~(\ref{eq-delta_SP}), the characteristic 
scale for collisionless reconnection layer changes. In particular, according 
to the published literature (e.g., Kleva~et~al.\ 1995; Rogers~et~al.\ 2001;
Cassak~et~al.\ 2007), 
the layer thickness for Hall collisionless reconnection becomes the ion 
acoustic Larmor radius, $\rho_s \sim c_s/\Omega_i$, where $\Omega_i$ 
corresponds to the total upstream magnetic field and where $c_s\sim 
(T_e/m_i)^{1/2}$ is the ion-acoustic speed. This result has been recently 
confirmed experimentally (Egedal~et~al. 2007). Then, the criterion for 
the onset of fast reconnection in the Hall-MHD regime becomes (e.g., 
Cassak~et~al.\ 2007)
\beq
\delta_{\rm SP} < \rho_s  \, .
\label{eq-condition-guide}
\eeq

According to the general scheme presented in the preceding section, 
we first need to consider the Sweet-Parker layer. Again, we want to 
estimate the value of the Spitzer resistivity at the center of the 
reconnection layer. Therefore, we need to determine the central electron 
temperature. One of the difficulties here is that, if there is a guide 
field, then the cross-layer force-balance in the form of equation~(\ref
{eq-pressure-balance}) is no longer applicable. This is because the 
pressure of the reconnecting component of the outside magnetic field, 
$B_0^2/8\pi$, no longer has to be balanced at the center of the layer 
by the plasma pressure alone. It can be partially balanced by an increase 
in the guide field pressure due to a modest compression.
In particular, a very strong guide field, $B_z \gg B_0$, 
dominates the pressure both inside and outside of the current 
layer and can be changed only slightly. Therefore, it essentially 
ensures incompressibility: $n^{\rm in}\simeq n^{\rm out}$. The temperature 
then decouples from the pressure balance and has to be determined
from some other considerations, namely, from the energy equation. 
Fortunately, most of the analysis presented in the previous section 
can be carried over to the strong-guide-field case without much change.
In particular, the plasma experiences Joule heating as it travels 
through the layer and, as long as one is in the collisional regime,
the parallel electron thermal conduction is not able to cool the
layer by a large factor. Therefore, we expect the temperature inside 
the layer to increase roughly to the equipartition level given by 
equation~(\ref{eq-T_e}). However, this equation can no longer be
regarded as equality, but just as a rough estimate. For the purposes
of this paper, though, this is good enough.

%
%
%
%
%

Now, let us discuss the right-hand side of condition~(\ref
{eq-condition-guide}). If the guide field is not very large 
compared with the reconnecting field, then one expects $\rho_s\sim d_i$, 
and hence the criterion~(\ref{eq-condition-1}) is not changed substantially. 
However, in the strong-guide-field case, we expect to have $\rho_s \ll d_i$ 
(assuming, as we always do throughout this paper, that upstream~$\beta\ll 1$).
Then, the criterion for transition to fast collisionless Hall reconnection 
is modified  structurally. It now becomes
\beq
\delta_{\rm SP} < \rho_s \sim
d_i\ \biggl( {{8\pi n_e T_e}\over{B_z^2}}\biggr)^{1/2} 
\sim d_i \  \beta_e^{1/2} \  {B_0\over{B_z}} \, ,
\eeq
where $\beta_e$ is based on the central electron temperature~$T_e$ 
and density and on the upstream reconnecting field component~$B_0$.
Note that, according to logic of our paper, this $\beta_e$ is to be 
estimated in the collisional regime. Nevertheless, it may also be 
useful to get some feeling for what $\beta_e$ (and hence~$\rho_s$)
should be immediately after the transition to fast reconnection. 
As we noted above, a strong guide field makes the plasma nearly 
incompressible, so that~$n_e$ is roughly uniform. The question 
then is what one should take for the electron temperature inside 
the layer? There is a considerable uncertainty here. In particular, 
there is no good reason for~$T_i$ to be close to~$T_e$ in the 
collisionless regime. For example, Hsu~et~al. (2001) present 
experimental evidence for strong ion heating $T_i\gg T_e$ inside 
a collisionless reconnection layer in the no-guide-field case in the~MRX, 
although the rise in~$T_i$ is considerably smaller in the presence of 
a guide field. In any case, with all these reservations, the simplest 
approach one can adopt is to take~$\beta_e\sim 1$ in the above expression. 
Then, the condition for fast collisionless reconnection in the strong guide 
field case becomes
\beq
{{\delta_{SP}}\over{\rho_s}} \sim
\biggl( {L\over{\lambda_{e,\rm mfp}}}\biggr)^{1/2} \ 
\biggl({m_e\over{m_i}}\biggr)^{1/4}\ {B_z\over{B_0}} < 1 \, ,
\eeq
which, after our usual manipulations, can be rewritten as
\begin{eqnarray}
L < L_c(B_z) &=& \sqrt{m_i\over{m_e}}\ \lambda_{e,\rm mfp} \ 
\biggl({B_0\over{B_z}}\biggr)^{2} \nonumber \\ 
&\simeq&  6\cdot 10^{9}\,{\rm cm}\, n_{10}^{-3}\, 
B_{1.5}^4\, \biggl({B_0\over{B_z}}\biggr)^{2} \,  
\label{eq-L_c-guide}
\end{eqnarray}
Thus, the critical global length $L_c$ in this case becomes smaller by 
a factor $(B_z/B_0)^2$ and so the collisionless reconnection condition 
is harder to satisfy. Also note that, for fixed~$n_e$ and~$B_z$, $L_c$ 
becomes very sensitive to the reconnecting field component: $L_c\sim B_0^6$.


\subsection{Caveats and Alternatives}
\label{subsec-caveats}

With all the above said, there are still many caveats and
alternative ideas that may potentially alter the above picture. 
It is important to be aware of them and try to gage their 
effect on the existence and the form of the fast collisionless
reconnection condition. I will discuss some of them in this
section.

First, there are a number of ideas that may, if correct, represent 
a potential challenge to our Conclusion~II (that collisional resistive-MHD 
reconnection is slow). For example, Lazarian \& Vishniac (1999) suggested 
that reconnection may be relatively rapid in pure resistive MHD in the 
presence of externally imposed three-dimensional MHD turbulence (see, 
however, Kim \& Diamond~2001 for a dissenting view). As far as I know, 
this idea has not yet been tested in three-dimensional (3D) numerical 
simulations, although 
it is definitely an interesting possibility that deserves further study. 
Such studies are indeed already underway (Vishniac~ 2007, private 
communication). Furthermore, even in the two-dimensional case, resistive-MHD 
Sweet--Parker-like current layer may become unstable to tearing 
instability when its aspect ratio exceeds a certain large number 
(probably a hundred) (Bulanov et al.\ 1978; Forbes \& Priest 1983; 
Biskamp~1986; Lee \& Fu 1986; Malyshkin~2005, private communication; 
Loureiro~et~al.\ 2007). Tearing may effectively result in an enhanced 
hyper-resistivity that would dominate over the normal resistivity (e.g., 
Strauss~1988). What will be the effect of the nonlinear development of 
this instability on the overall reconnection rate is still not clear.
In addition, Dahlburg~et~al.\ (1992) identified a mechanism, based on 
a secondary ideal-MHD instability, that leads to the excitation of fully 
three-dimensional turbulence within a reconnecting current layer. 
Finally, in a recent numerical simulation with both uniform resistivity 
and Hall effect, Cassak~et~al.\ (2005) observed an interesting hysteresis 
phenomenon: as the resistivity was turned down, there was a transition 
from slow Sweet--Parker to fast Hall reconnection, in accordance with 
condition~(\ref{eq-condition-1}).
However, as the resistivity was turned back up to a level
higher than that corresponding to~(\ref{eq-condition-1}), 
the system still stayed in the fast mode. 
This puzzling behavior needs to be investigated further.

Second, there are recent studies relevant to Conclusion~III
(that collisionless reconnection is fast). In particular, 
even though most of the authors agree that Hall reconnection 
is much faster than Sweet--Parker, there is still a significant 
disagreement about exactly how fast it actually is. For example, 
Drake and his collaborators report a universal reconnection rate 
of about~0.1~$V_A$, independent of the electron microphysics that 
is ultimately responsible for breaking the field lines (e.g., 
Biskamp~et~al.\ 1997; Shay~et~al.\ 1998, 1999, 2001; Cassak~et~al. 2005; 
see also Birn~et~al.\ 2001). On the other hand, others argue that 
it should not be universal but should instead be a function of the 
system parameters, such as the electron/ion mass ratio (e.g., 
Bhattacharjee et al. 2001). 
In addition, Daughton~et~al.\ (2006) recently raised a concern about 
the validity of periodic boundary conditions used in the majority of 
previous numerical studies. In particular, they argued that the reported 
very high reconnection rate may be an artifact of such boundary conditions. 
Furthermore, they performed 2D particle simulations with the appropriate 
open boundary conditions and observed that, after an initial transient peak 
consistent with the high value reported by Shay~et~al. (1999, 2001), the 
reconnection rate went down and settled at a significantly lower level 
(Daughton~et~al. 2006). Similar conclusions were reached independently 
by Fujimoto (2006) using periodic 2D particle simulations but in a very 
large. After a year-long debate, however, this controversy finally seems 
to have been resolved. It is now agreed by both sides that the electron 
dissipation region develops a two-scale structure with an extended 
high-velocity electron jet in the outflow direction; however, the 
dimensionless reconnection rate still remains high  ($v_{\rm rec}/V_A 
\gg d_i/L$), although perhaps not quite as high as previously believed 
(e.g., Shay et al. 2007; Karimabadi~et~al.\ 2007).

Finally, I would like to mention that the picture presented
in the above two sections, implies certain physical conditions
and thus is not expected to be universally valid in all astrophysical
environments. For example, in the case of weakly-ionized plasma in 
molecular clouds (e.g., in the context of star formation),
reconnection dynamics is strongly modified by recombination
processes and may become significantly faster than Sweet--Parker 
even in the collisional regime (Zweibel~1989; Heitsch \& Zweibel~2003).
Also, in the case of reconnection of super-strong magnetic fields
in magnetar magnetospheres ($B_0\gtrsim 10^{14}$~G), the density 
of the released magnetic energy is so large that the pressure 
inside the layer is dominated by radiation. The resulting temperature 
is so high that a large number of electron-positron pairs is produced;
this means that the number of particles  is not conserved, and so the 
above estimates would have to be modified (see, e.g., Thompson~1994; 
Lyutikov \& Uzdensky 2003; Uzdensky \& MacFadyen 2006).

Other issues that are worth-mentioning and that may substantially 
complicate the simple physical picture presented in this paper are: 
impulsive, bursty reconnection (e.g., Bhattacharjee~et~al. 1999;
Bhattacharjee~2001; Daughton~et~al.\ 2006, 2007, in preparation); 
and reconnection in three dimensions (e.g., Longcope~1996).


\section{Solar Coronal Heating as a Self-Regulated Process}
\label{sec-corona}


\subsection{General Picture}
\label{subsec-corona-general}

The physical picture presented in this paper has a number 
of interesting implications for the solar corona that I will 
discuss in this section.

At the present, the theory of magnetic reconnection has not yet reached 
the level of maturity that would allow us to make accurate predictions 
regarding when the transition between the slow and fast reconnection modes 
will occur. 
Moreover, there is still no consensus on the exact nature of the fast 
collisionless reconnection mechanism and on the actual scalings of the 
reconnection rate in either fast or slow reconnection (see section~\S~\ref
{subsec-caveats} for more discussion). However, what is important for us in 
this paper is just the sole fact of existence of such a transition and that 
it has to do with the plasma collisionality. The transition doesn't actually 
have to be razor-sharp, and in reality there may be not two but more different 
regimes with different reconnection rates and hence more than one such 
transition. Also, the transition condition may depend on other physical 
parameters, such as the presence of a guide field (see \S~\ref
{subsec-guide-field}) or, perhaps, the effectiveness of radiative 
cooling, which we have omitted in this paper. Whatever the case, 
the existence of two reconnection regimes with a collisionality-related 
transition between them is a very important fact and we would like to 
exploit it in various astrophysical applications. In this section, 
I will apply this condition [namely, eqs.~(\ref{eq-condition-2}) 
and~(\ref{eq-L_c-final})] to the problem of solar coronal heating.

I propose that {\it coronal heating is a self-regulating process
that works to keep the corona marginally collisionless}, in the 
sense of equations~(\ref{eq-condition-1}) and~(\ref{eq-condition-2}) 
(see also Uzdensky~2006, 2007).

According to Parker's nanoflare theory (Parker 1972, 1983, 1988), 
as long as the twisting and braiding of coronal loops by photospheric
footpoint motions and flux emergence episodes keep producing thin current 
sheets in the corona (e.g., van~Ballegooijen~1986), magnetic dissipation 
in these current sheets leads to continuous coronal heating. In reality, 
this heating, of course, is not uniformly distributed; it is instead 
strongly intermittent, localized in both space and time (Rosner~et~al. 
1978; hereafter~RTV78). 
This basic picture of coronal heating has been shown to work in several 
3D MHD numerical simulations (e.g., Galsgaard \& Nordlund 1996; Hendrix 
et al. 1996; Gudiksen \& Nordlund 2002, 2005).

Now, the overall, integrated heating density, i.e., the rate of magnetic 
dissipation per unit volume, depends on the reconnection rate in these 
sheets, e.g., on whether reconnection is fast (e.g., Petschek) or slow 
(Sweet--Parker). This is actually a rather subtle point. Indeed, in a 
steady state, the energy dissipated in the corona per unit time should 
be equal to the power pumped into the corona magnetically from the solar 
surface. 
And it is not obvious why and how the energy-pumping rate depends on 
what happens in the corona. For example, if the corona were enclosed 
in a fixed volume, then the energy dissipation per unit volume would 
be fixed (as in driven MHD turbulence in a box). However, it is important to
recognize that the coronal volume is not fixed! If, for example,  
reconnection were to suddenly slow down, then more energy would be 
pumped in than the corona could dissipate; then, in order to accommodate 
this additional free magnetic energy, the corona would respond by  
increasing its scale-height. That is, coronal magnetic structures would 
grow in height until, finally, the total dissipation in the corona became 
equal to the total input from the photosphere. Because of the increased 
volume, the magnetic dissipation per unit volume is decreased.

In addition, as coronal magnetic structures grow in height, 
especially when $H\gtrsim R_\odot$, the amount of energy pumped 
into the corona by the footpoint motions may go down. 
This is because the work done by footpoint motions is 
proportional to ${\bf v}_{\rm fp}\cdot {\bf B}_{\rm hor} B_{\rm vert}/4\pi$. 
As the coronal structures grow in height without increasing their lateral 
size, the horizontal field, $B_{\rm hor}$, decreases, whereas the vertical 
field component, $B_{\rm vert}$, does not change. Correspondingly, 
the overall power pumped from the photosphere into the corona goes down.

As follows from equation~(\ref{eq-L_c-final}), the reconnection mode
is determined by the global scale~$L$ of the reconnection layer and by 
the basic physical parameters characterizing the plasma in the layer 
(i.e, $n_e$, $T_e$, and~$B_0$). The typical values of~$L$ and~$B_0$ 
are determined by the scale and strength of the magnetic structures 
emerging from the Sun and by the scale of footpoint motions (e.g., 
the meso-granular scale). Therefore, for the purposes of the present
discussion, I will regard~$L$ and~$B_0$ as fixed and then ask what 
determines the coronal density and temperature. 

Following this line of reasoning, let us invert equation~(\ref{eq-L_c-final})
and view it as the condition for the plasma density. That is, let us 
introduce a scale-dependent critical density, $n_c$, below which the 
reconnection process transitions from the slow collisional Sweet--Parker 
regime to a fast collisionless regime:
\beq
n_c (B_z\lesssim B_0) 
\sim 2\cdot 10^{10}\, {\rm cm}^{-3}\, B_{1.5}^{4/3}\, L_9^{-1/3} \, ,
\label{eq-n_c}
\eeq
where $L_9$ is the global reconnection layer length expressed 
in units of~$10^4$~km. In the strong-guide-field case, $B_z\gg B_0$,
the corresponding expression becomes:
\beq
n_c (B_z \gg B_0) 
\sim 2\cdot 10^{10}\, {\rm cm}^{-3}\, B_{1.5}^{4/3}\, L_9^{-1/3} 
\biggl({{B_0}\over{B_z}}\biggr)^{2/3} \, .
\label{eq-n_c-guide}
\eeq
It is interesting to note that these values are close to those observed 
in active solar corona. I suggest that this is not a pure coincidence.

The reason for this is that there is an important feedback between coronal 
energy release and the coronal density. This feedback is due to the fact 
that the gas high in the corona actually comes from evaporation from the 
surface along the field lines that just underwent reconnection. In fact, 
it is essential to consider the coronal heating not only as the process 
of increasing the coronal plasma's temperature, but also as that of 
increasing its density (e.g., Klimchuk~2006; Aschwanden~et~al. 2007).
Indeed, a part of the energy released in a reconnection event is rapidly 
conducted along the field by fast (perhaps, nonthermal) electrons to the 
surface and is deposited in a denser photospheric and chromospheric plasma. 
This leads to a localized heating at the footpoints of the post-reconnected 
magnetic loops and to a subsequent chromospheric evaporation along them, 
a well-documented phenomenon in solar corona (e.g., Aschwanden~et~al.\ 2001a,b 
and Klimchuk~2006; see, however, Brown~et~al. 2000 and Aschwanden~et~al.\ 2007 
for an evidence that chromospheric evaporation occurs in response to 
chromospheric, rather than coronal, heating events).
As a result, these loops become filled with a dense and hot plasma. 
The density rises and may now exceed~$n_c$. This will shut off any
further reconnection (and hence heating) involving these post-reconnective
loops until they again cool down, which occurs on a longer, radiative 
timescale.

Let us consider a couple of examples of how this might work.
First, let us consider the case when the plasma density is 
relatively high and so the radiative cooling time, $\tau_{\rm rad}$, 
is relatively short compared with the timescale on which the coronal 
magnetic structures change due to footpoint motions, $\tau_{\rm fp}$; 
however, it is of course still long compared with the fast reconnection 
timescale. The opposite limiting case will be discussed later in this section.
Let us suppose that due to field-line twisting, a reconnecting structure 
is set up in the corona at $t=0$. Let this structure be characterized by 
the current sheet length~$L$ and the reconnecting field component~$B_0$. 
Furthermore, let us further suppose that, initially, the ambient density 
of the background plasma is higher than~$n_c(L,B_0)$. Then the reconnection 
layer is collisional and reconnection proceeds very slowly, in the 
Sweet--Parker mode. That is, there is almost no reconnection at all 
and hence coronal heating is weak. The surrounding plasma gradually 
cools by radiation (and also, in general, by thermal conduction) and 
the pressure scale height gradually goes down. 
The gas gradually precipitates to the surface. 
Then the density of the plasma entering the layer decreases and at some 
point becomes lower than the critical density. The reconnection process 
then suddenly switches to the collisionless regime. Petschek-like fast 
reconnection ensues, and the rate of magnetic energy dissipation greatly 
increases. A flare commences. Some fraction of the energy released by 
reconnection is transported by the electron conduction along the reconnected 
field lines down to the base, where it is deposited in the much denser 
photospheric or chromospheric plasma. This in turn leads to a massive 
evaporation along the same field lines. As a result, the newly-reconnected 
loops are now populated with relatively dense and hot plasma. They cool down 
only slowly via conduction and radiation losses, keeping their relatively high 
density for an appreciable length of time. If, during this time, these loops 
become twisted or somehow get in contact with other loops, they are now not 
likely to reconnect rapidly, since their plasma density is above critical. 
This inhibits further coronal heating in the given region. In fact, we can 
speculate that for any further outbursts of coronal activity in the given 
region to occur, one has to wait for the gas in post-reconnective loops to 
cool down significantly, which occurs on a longer, radiative timescale.

Thus we see that, although highly intermittent and inhomogeneous, 
the corona is working to keep itself roughly at about the height-dependent 
critical density given by equation~(\ref{eq-n_c}) or~(\ref{eq-n_c-guide}). 
Correspondingly, the background coronal temperature should be such that 
results in a density scale-height that is just large enough to populate 
the corona up to the critical density level at a given height. 
In this sense, coronal heating regulates itself (Uzdensky~2006, 2007).

An important remark is that loop brightness in soft x-rays 
and UV does not necessarily mean that this loop is actively undergoing 
fast reconnection at the given moment of time. Instead, it just  
reflects the fact that the loop has a high plasma density (since radiation 
intensity is proportional to~$n^2$). This implies an episode of strong 
evaporation from the surface in the recent past, presumably caused by a 
{\it preceding} fast-reconnection coronal energy release in this loop. 
The nearby coronal regions that are darker have evidently lower density, 
and they would be ripe for reconnection from the point of view of the 
collisionless reconnection condition. However, they probably are filled 
with a magnetic field that does not have complex topology with current 
sheets, since the underlying  reason for coronal activity 
is always the generation of small coronal current sheets as a result of 
field-line braiding driven by a complicated pattern of the footpoint 
motions on the solar surface.

The above radiative-cooling-dominated scenario was presented here first 
only because it brings out more clearly the role that plasma density 
plays in controlling the reconnection regime. It also highlights the 
role of chromospheric evaporation in controlling the plasma density. 
However, there is an alternative version of this picture corresponding 
to the opposite situation, $\tau_{\rm fp}<\tau_{\rm rad}$. 
In this case one can repeat the same arguments as those given above but 
instead of the slow initial evolution of the plasma density due to gradual 
radiative cooling, one can invoke the slow evolution of the reconnecting 
magnetic field strength caused by the motion of the loop footpoints (as
was described by Cassak et al. 2006). 
In other words, one can regard the ambient plasma density~$n$ as fixed 
on the timescale of interest, whereas the reconnecting field component~$B_0$ 
gradually increases (for simplicity we also keep~$L$ fixed). Then, the
current sheet gradually builds up, while staying in the Sweet--Parker 
mode, and then rapidly switches over to the fast collisionless mode 
once~$B_0$ has exceeded a certain threshold. This scenario is in fact 
closer in spirit to the original Parker nanoflare model (Parker 1983, 1988), 
and also to the discussion by Cassak~et~al.\ (2005, 2006).
Instead of defining $n_c(L,B_0)$, one can just as well rewrite the fast 
reconnection condition in terms of a critical current-sheet strength, i.e.,
the critical reconnecting magnetic field component, $B_c$, expressed as
a function of~$L$ and~$n$:
\beq
B_c(n,L) \simeq 20\ {\rm G}\ L_9^{1/4}\ n_{10}^{3/4} \, .
\label{eq-B_c}
\eeq
In the case of a strong guide field, $B_z\gg B_0$, the critical 
reconnecting field is
\beq
B_c(n,L,B_z) \simeq 30\ {\rm G}\ L_9^{1/6}\ n_{10}^{1/2} \ B_{z,2}^{1/3} \, .
\label{eq-B_c-guide}
\eeq
where $B_{z,2}\equiv B_z/(100\, {\rm G})$.

As I will show below, in the long run one can expect some sort of statistical 
steady state with $\tau_{\rm fp}\sim\tau_{\rm rad}$. Then, this second 
scenario is just as likely to take place as the first one. In fact, in 
reality they probably both occur alongside each other.

I would like to emphasize that the phenomenon described in the first part of 
this paper (i.e., the transition from slow to fast collisionless reconnection) 
is not the whole coronal heating story, although, in my view, it is an 
important part of it. Conceptually, it is probably about one quarter of 
the whole story. The other important pieces of the puzzle are:
(i) complex dynamic interaction of a large number of coronal magnetic 
loops, ultimately responsible for the generation of current sheets in 
the corona by footpoint motions; this requires the development of an
appropriate statistical description of loops (e.g., Uzdensky \& Goodman 
2007, in preparation);
(ii) chromospheric evaporation following a reconnection event
and motions of plasma along the loop;
and 
(iii) radiative (and thermal-conduction) energy losses and the resulting 
precipitation of the material back to the solar surface.
All these essential physics ingredients (or at least some effective 
representation of them) will eventually have to be incorporated into
a single comprehensive theoretical (probably numerical) model of the 
solar corona.

It is important to note that each of these processes comes 
with its own characteristic timescale. The reconnection time 
in the fast regime, $\tau_{\rm rec}$, is very short. For the 
purposes of our discussion, we shall regard it as instantaneous. 
The characteristic time for material evaporation following a 
nanoflare is also relatively short. The two longer time-scales 
in the problem are the characteristic timescale for generation 
and growth of new current sheets by photospheric footpoint
motions, $\tau_{\rm fp}$, and the radiative cooling timescale
$\tau_{\rm rad}$. As we shall show below, the system tends to 
adjust itself to a statistical equilibrium in which these two 
timescales are comparable (of order one hour in the solar corona). 
Finally, the longest time scale of all is the Sweet--Parker reconnection 
time. It is much longer than one hour, and is therefore irrelevant in 
our problem. That is, the energy release during the Sweet--Parker 
phase is not important. However, this does not mean that the existence 
of the slow Sweet--Parker regime is not important; it is in fact 
extremely important, since it allows for a strong current sheet 
to build up, accumulating a large amount of free magnetic energy. 

This discussion brings us to one of the most interesting issues in 
solar coronal heating --- intermittency. The key questions here are: 
what is the distribution of (spatial, temporal, and energy) scales 
on which magnetic dissipation occurs? what scales are responsible 
for most of the energy release? In my opinion, the collisionless 
reconnection condition is, at the end of the day, very important 
for answering these questions. In particular, I think this condition
plays a key role in establishing the spectrum of the energy release 
events, e.g., the observed power-law distribution of flare energies.


\subsection{Long-term Evolution of a Coronal Loop}
\label{subsec-corona-loop}

To illustrate the above points, let us consider the following model problem.
One can raise the question of whether the self-regulating behavior of 
the nonlinear dynamical system described above can exhibit an oscillatory 
behavior (Parker 2006, private communication), especially in light of a 
hysteresis-like phenomenon reported by Cassak et al. (2005).
Of course, in general this issue is difficult to assess within the present 
model, since it first needs to be incorporated in the overall multi-loop 
dynamics (e.g., Uzdensky  \& Goodman 2007, in preparation); that is, the 
energy release in a given fast reconnection event does not, strictly speaking, 
directly affect the reconnection process that produced it. 
However, it does lead both to relaxation of the magnetic stress and also, 
importantly, to the increase in density in the post-reconnected flux tube 
(or more precisely, in the elementary magnetic thread within the larger 
flux tube). This will in turn temporarily inhibit any reconnection 
involving this given field thread even if a new current sheet is formed. 
This inhibition will last for some time until either the plasma density 
goes down because of the radiative cooling or the reconnecting field 
component in the new current sheet grows to a larger strength.

Let us consider a simple illustrative example that demonstrates 
this oscillatory behavior and the secular evolution of the system
on a longer timescale.
Consider a composite coronal flux tube, initially filled with a relatively low 
plasma density~$n$. For simplicity, let us imagine that that this tube consists
of two separate elementary flux threads that are being wrapped around each 
other steadily by the sheared photospheric motions. Consequently, a current 
sheet gradually builds up inside the tube, in the spirit of the nanoflare 
model. Again for simplicity, let us assume that the length~$L$ of this current 
sheet and the axial magnetic field strength $B_z$ in the tube stay constant at 
all times. Then, as the elemental flux threads are wrapped around each other,
the reconnecting field component, $B_0$, gradually increases. 
Because the plasma density is low, the radiative cooling time is relatively
long and therefore we can regard the plasma density constant during this phase.
On the other hand, also because the density is low, the critical field strength
is also low, according to equation~(\ref{eq-B_c-guide}). That is, the current 
sheet will become collisionless relatively early, at some small value of~$B_0$.
Correspondingly, the amount of free magnetic energy attributed to the current 
sheet at that moment will also be small and the resulting reconnection event 
will be relatively weak. After the reconnection event, the magnetic field
becomes unwrapped again (almost completely, since the hysteresis behavior 
reported by Cassak et al. 2005, 2007 implies that fast reconnection proceeds 
to the end); then the footpoint-driven process of current-sheet build-up starts
all over again. Now, if the plasma density in the tube under consideration 
were fixed at a constant small value~$n_e$, then (since $L$ and~$B_z$ are 
also kept fixed) all the subsequent fast reconnection events would be 
triggered at the same critical value of~$B_0=B_c(n,L,B_z)$. 
Correspondingly, they would all be characterized by a relatively small 
amount of energy released, on the order of~$V B_c^2(n_e,L,B_z)/8\pi$, 
where~$V$ is the volume of the  region involved in the given reconnection 
event. At the same time, since weaker current sheets are easier to produce, 
there would be following each other at a relatively rapid rate.

However, because of the mass exchange between the surface of the Sun and 
the corona, the ambient density is not fixed a priori. In particular, if 
one starts with a small initial density, the first few reconnection events 
will be energetically small, but, nevertheless, they will eventually lead 
to an increase in the plasma density via chromospheric evaporation. 
This will make the critical magnetic field stronger and hence the amount 
of free energy stored between reconnection events and released through 
them larger. At the same time, such events will be more rare, since it 
takes longer to build up a stronger current sheet. Therefore, an increase 
in plasma density shifts the dominant  energy-release scale towards larger 
energies. Thus, in this picture, the plasma density in the tube increased 
over time in a step-like manner: it increases by a certain amount in the 
aftermath of each reconnection event and stays roughly flat between them. 
The amount of plasma pumped into the corona as a result of each event is 
roughly speaking proportional to the magnetic energy released in this event. 
The latter is in turn a monotonically-increasing function of density, 
according to equation~(\ref{eq-B_c-guide}). That is, we have, keeping~$L$ 
and~$B_z$ fixed: $\delta n \sim B_c^2(n) \sim n$. Thus, the relative 
increase is then independent of density; it can be roughly estimated 
as follows. Consider a loop of constant cross-sectional area~$L^2$ and 
length~$L_\parallel>L$; the loop's volume is then $V_{\rm loop}\sim 
L_\parallel\, L^2$. Assuming that the reconnection layer extends all 
the way along the loop, the volume~$V$ of the region involved in the 
reconnection event is similar: $V\sim V_{\rm loop} \sim L_\parallel\, L^2$. 
Then, the magnetic energy released in the event is of order $E_{\rm rec}\sim 
(B_0^2/8\pi)\, L_\parallel\, L^2$. Next, let us say that a certain fraction 
$\epsilon<1$ of this energy goes towards filling the entire loop with plasma 
evaporated from the surface. The coefficient $\epsilon$ depends on the portion 
of the released energy that goes to the energetic electrons and is transported 
by them to the photosphere, times the fraction of this deposited energy that 
goes to material evaporation (as opposed to that immediately radiated away 
from the footpoints). Overall, $\epsilon$ describes the fraction of the flare 
energy that eventually will be radiated by post-coronal loops as soft X-rays, 
UV, etc., but not hard X-rays. Immediately after the evaporation episode, 
the energy $\epsilon E_{\rm rec}$ is divided between the increase in the 
gravitational energy, $\delta E_{\rm grav}$, and in the thermal energy, 
$\delta E_{\rm th} \simeq 3\, \delta n \, k_B T_{\rm loop} V_{\rm loop}$, 
of the coronal-loop plasma (for simplicity of discussion, I dropped the 
contribution to~$\delta E_{\rm th}$ associated with the increase in the 
gas temperature; in reality, of course, both~$n$ and~$T_{\rm loop}$ should 
increase together). If the height of the loop is larger than the density 
scale-height, then these two contributions are likely to be comparable, 
according to the virial theorem. However, in the case of a compact loop, 
whose height is less than the thermal scale-height, the thermal energy 
increase will be larger than~$\delta E_{\rm grav}$. In either case, we 
expect $\delta E_{\rm th} \sim \epsilon E_{\rm rec}$, from which we 
immediately find
\beq
{{\delta n}\over{n}} \sim 
\epsilon\, {{B_c^2}\over{8\pi}}\, {1\over{3 n k_B T_{\rm loop}}} =
{2\epsilon\over{3\,\beta_{\rm loop}}} \, ,
\label{eq-delta-n-1}
\eeq
where $\beta_{\rm loop} \equiv 16\pi\, n k_B T_{\rm loop}/B_c^2$
is the characteristic plasma-beta based on the critical reconnecting 
field. Substituting the expression~(\ref{eq-B_c-guide}) for $B_c$, we 
get
\beq
\beta_{\rm loop} \simeq 0.07\, n_{10}\, T_{\rm loop,6}\, B_{c,1.5}^{-2} \sim
0.07\, T_{\rm loop,6}\, B_{z,2}^{-2/3}\, L_9^{-1/3} \, ,
\eeq
and so 
\beq
{{\delta n}\over{n}} \sim 
10\, \epsilon \, T_{\rm loop,6}^{-1}\, B_{z,2}^{2/3}\, L_9^{1/3} \, .
\label{eq-delta-n-2}
\eeq
There are also numerical factors of order unity which we have ignored.
It is reasonable to expect that this ratio should be roughly of order 
unity. For now let us just assume that it is not large, i.e., 
$\delta n/n \lesssim 1$. 

As mentioned above, in a realistic scenario the loop temperature should 
also increase at each step, in concert with the density. This means that,
to get into the coronal loop, chromospheric gas would need to be heated 
to a temperature that is higher than at the previous step. Then, each 
erg of the released energy would be able to lift a smaller amount of gas, 
i.e., the evaporation efficiency of coronal heating would decrease over 
time. 
To illustrate this point, let us, for example, adopt the famous RTV scaling 
for the loop temperature (RTV78), even though it was derived assuming steady 
heating and thus is not valid for the impulsive situation considered here 
(see below for more discussion). Recasting the RTV78 scaling in terms of 
the electron density instead of the pressure, we write $T_{\rm loop}^{\rm RTV} 
\simeq 3\cdot 10^6\ {\rm K}\ n_{10}^{1/2}\, L_{\parallel,9}^{1/2}$. 
Substituting this into equation~(\ref{eq-delta-n-2}), we get
\beq
{{\delta n}\over{n}} \sim 3\, \epsilon\, n_{10}^{-1/2}\, 
B_{z,2}^{2/3}\, L_9^{1/3}\, L_{\parallel,9}^{-1/2} \sim n^{-1/2}\, .
\label{eq-delta-n-RTV}
\eeq

Next, the time separation between the steps also increases with the 
increased threshold magnetic field. In particular, let us assume that 
the footpoint driving twists up the field lines and thus creates and 
then strengthens current sheets (or perhaps, quasi-separatrix layers, 
see, e.g., Galsgaard \& Nordlund 1996; Titov \& Hornig 2002) at some 
constant rate. This is, of course, an over-simplification. In reality, 
current sheets may be created as a consequence of the nonlinear development 
of kink-like instabilities of twisted magnetic loops (Dahlburg~et~al. 2005). 
Then, the growth of~$B_0$ with time is not likely to be a smooth function, 
but instead may involve sudden jumps. Here, however, we would like to avoid 
such complications. Thus, let us consider, as an illustration, a simple model
in which the reconnecting field component increases linearly between 
reconnection events,
\beq
{dB_0\over{dt}} = \gamma B_z \, ,
\eeq
where $\gamma = {\rm const}$. Then, using equation~(\ref{eq-B_c-guide}), 
the time~$\delta t$ it takes the reconnecting field to grow a given critical 
value~$B_c(n)$ (after a complete relaxation during the preceding reconnection 
event) is $\delta t = \gamma^{-1} B_c(n)/B_z \sim n^{1/2}$. Therefore, as long 
as the relative increase in density at each step is not large, the long-term 
($t\gg \delta t$) evolution of the system can approximately be described by 
a differential equation:
\beq
{dn\over{dt}} \sim {{\delta n(n)}\over{\delta t(n)}} \sim {\rm const} \, ,
\eeq
corresponding to a linear rise $n(t)\sim t$. Accordingly, 
the emission measure of the loop should increase as~$t^2$. 
[Alternatively, if we neglect the evolution of the loop temperature 
in the above analysis and instead keep $T_{\rm loop}={\rm constant}$, 
we then get $\delta n\sim n$ and hence $n(t)\sim t^2$, with the emission 
measure rising as~$t^4$.]

This growth will continue until one of the following two processes will 
intervene. First, it may happen, as the density builds up, the critical 
value of~$B_0$ will become so large (e.g., a sizable fraction of~$B_z$), 
that the equilibrium shape of the entire loop will be affected. 
For example, the loop may become external-kink-unstable and become helical 
(sigmoidal loop), which with further twisting may result in a large-scale 
disruption with a catastrophic energy release, a large flare. 
This scenario, by the way, suggests that coronal loops should gradually 
(on the time scale of hours) brighten up before a major disruption, 
because a higher plasma density (manifested as a higher emission measure) 
leads to a higher critical field strength and hence allows a larger amount 
of free magnetic energy to be accumulated without being prematurely dissipated 
via smaller reconnection events.

On the other hand, there is another, less violent outcome that is also 
possible. Indeed, so far we have been neglecting radiative cooling. 
However, as the plasma density gradually builds up due to a series 
of chromospheric evaporations caused by reconnection events, the 
radiative cooling time decreases (as $n^{-1}$). At the same time, 
as we have seen above, the time interval between subsequent reconnections 
becomes longer and longer. Eventually, a point may be reached when the amount 
of plasma precipitated during $\delta t$ will become equal to the amount of 
material injected into the loop during an evaporation episode. 
After that point, there will be no further net secular change 
in the coronal density; the system will attain a stable limit
cycle behavior. To evaluate the exact conditions characterizing this 
state one will need to calculate the amount of material evaporated
following a given small flare and the rate at which the plasma 
returns to the photosphere as a result of gradual cooling. 
This would require a detailed model of the thermal structure along loop 
(e.g., along the lines of RTV78), but this can (and should!) 
definitely be done. For now, however, we just want to get some qualitative 
feeling. Therefore, we shall just say, tentatively, that the system 
enters this stable cyclic regime (with no secular density gain) when 
the radiative cooling time becomes equal to the time between reconnection 
events. This yields the following condition:
\beq
\delta t (n) = \gamma^{-1}\ {{B_c(n)}\over{B_z}} = \tau_{\rm rad} (n) \, .
\label{eq-limit-cycle-cond}
\eeq

This equation can now be viewed as a condition that determines 
the long-term equilibrium plasma density inside the loop, $n_*$. 
Substituting equation~(\ref{eq-tau_rad}) for the radiative cooling 
time and equation~(\ref{eq-B_c-guide}) for~$B_c$, we obtain~$n_*$
as a function of the loop temperature~$T_{\rm loop}$, $\gamma$, $L$, 
and~$B_z$:
\beq
n_* \simeq 
2\cdot 10^{10}\, {\rm cm}^{-3}\ B_{z,2}^{4/9}\, L_{9}^{-1/9}\, 
(\gamma\tau_{\rm rad,0})^{2/3} \,
\biggl({{T_{\rm loop,6}}\over{Q_{-22}}}\biggr)^{2/3}\, ,
\label{eq-n*}
\eeq
where $Q_{-22}\equiv 
Q(T_{\rm loop})/(10^{-22}\,{\rm erg}\,{\rm cm}^3\,{\rm s}^{-1})$ 
and where for convenience we defined $\tau_{\rm rad,0}$ as the 
radiative cooling time corresponding to $n_{10}=1$, 
$T_{\rm loop}=1\,{\rm MK}$, and~$Q_{-22}=1$: 
$\tau_{\rm rad,0} \simeq 400\ {\rm sec}$.

Correspondingly, the characteristic threshold reconnecting magnetic 
field is 
\begin{eqnarray}
B_{0,*} &=& B_c(n_*,L,B_z) \nonumber \\
&\simeq& 45\,{\rm G}\ L_9^{1/9}\, B_{z,2}^{5/9}\, 
(\gamma\tau_{\rm rad, 0})^{1/3}\, 
\biggl({{T_{\rm loop,6}}\over{Q_{-22}}}\biggr)^{1/3}\, ,
\label{eq-B_0*}
\end{eqnarray}
and the characteristic time interval between subsequent reconnection events 
is
\begin{eqnarray}
\Delta t_* &=& \tau_{\rm rad}(n_*) \nonumber \\
&\simeq& 200\, {\rm sec}\ B_{z,2}^{-4/9}\, L_{9}^{1/9}\,
(\gamma\tau_{\rm rad,0})^{-2/3} 
\biggl({{T_{\rm loop,6}}\over{Q_{-22}}}\biggr)^{1/3}\, ,
\label{eq-delta-t*}
\end{eqnarray}

The characteristic amount of energy released in each reconnection event 
comprising the limit cycle is 
\begin{eqnarray}
E_* &\simeq& L_\parallel\, L^2 \, {{B_{0,*}^2}\over{8\pi}} 
\simeq 8\cdot 10^{28}\, {\rm erg}\ \times \nonumber \\
&\times&\, L_{\parallel,9}\, L_9^{2+2/9}\, 
B_{z,2}^{10/9}\, (\gamma\tau_{\rm rad,0})^{2/3}\, 
\biggl({{T_{\rm loop,6}}\over{Q_{-22}}}\biggr)^{2/3}\, ,
\label{eq-E*}
\end{eqnarray}
and hence the time-averaged rate of magnetic dissipation in the loop 
is
\begin{eqnarray}
H_* &=& {{E_*}\over{\Delta t_*}} 
\simeq 4 \cdot 10^{26}\, {\rm erg}\, {\rm sec}^{-1}\, \times \nonumber \\
&\times&\, L_{\parallel,9}\, L_9^{2+1/9}\, B_{z,2}^{14/9}\, 
(\gamma\tau_{\rm rad,0})^{4/3}\, 
\biggl({{T_{\rm loop,6}}\over{Q_{-22}}}\biggr)^{1/3} ,
\label{eq-H*}
\end{eqnarray}

Note that most of the above parameters have weak dependence 
on~$L$, except for~$E_*$ and~$H_*$, whose strong $L$-dependence 
is almost entirely due to the volume involved.

To make any further progress, we need two more relationships: 
(1) the cooling function~$Q(T)$ and (2) a scaling for the loop 
temperature~$T_{\rm loop}$. Note that both are subject to 
significant uncertainties. In particular, the radiative cooling 
function exhibits a complicated behavior in the relevant temperature 
range (1-10~MK); for example, according to Cook~et~al. (1989), it 
decreases sharply between $T=1$~MK and $T=3$~MK, and then stays 
nearly flat for $T>3$~MK. Therefore, in principle, one should not 
expect simple power-law scalings to result at all. Furthermore, 
there are still significant disagreements in the literature 
regarding~$Q(T)$ (e.g., Raymond~et~al. 1976; Raymond \& Smith 1977; 
RTV78; Cook~et~al.\ 1989; Landi \& Landini 1999; Aschwanden~et~al.\ 2000a). 
For these reasons, any realistic quantitative analysis of radiative 
cooling in the context of the present model is a highly non-trivial 
task, adding even more ambiguity to this already complicated picture. 
Getting into such a high level of sophistication and detail would 
exceed the overall level of accuracy of our model. We shall therefore 
leave it for a future study. 

In addition to the cooling function, we need an expression for 
the loop temperature, $T_{\rm loop}$, which strongly affects plasma
cooling, in terms of the other loop parameters. In principle, the 
temperature, or, more generally, its distribution along the loop, 
is determined by the hydrostatic balance in conjunction with the 
energy transport along the loop, which includes thermal conduction, 
radiative losses, and distributed heating. Such an analysis has been 
first performed by RTV78 for the simplest case of uniform and steady 
heating; as a result, they derived the famous RTV scaling for the 
loop-top temperature: 
$T_{\rm loop}^{\rm RTV} = 1.4\cdot 10^3 (p L_\parallel)^{1/3}$, 
where $p$ is the loop pressure and~$L_{\parallel}$ is its length 
(which is different and usually larger than our current-sheet 
length~$L$; in our model, we shall regard it as constant). 
Now, it is true that there is a lot of disagreement in the modern 
literature regadring the applicability of the RTV scaling to the real 
solar corona and the observational support is doubtful (e.g., Porter 
\& Klimchuk 1995; Cargill~et~al.\ 1995; Kano \& Tsuneta 1995; 
Aschwanden~et~al.~2000b,2001a). This is not surprising taking into account 
that the RTV scaling was derived assuming stationary and uniform heating, 
and thus ignored the impulsive and intermittent nature of coronal energy 
dissipation (which is an intrinsic property of my model). In addition, 
to develop their theory, RTV78 relied on an old cooling function due to 
Raymond (see below), whereas somewhat different cooling functions have 
been used in recent years (Cook~et~al.\ 1989; Landi \& Landini 1999; 
Aschwanden~et~al.\ 2000a). 

Nevertheless, despite its significant limitations, the RTV scaling has 
been highly influential in solar physics. It is still a highly-respected 
common standard, against which solar physicists often measure their theories 
and observations. Therefore, just to present an illustrative example, let 
us now combine the RTV scaling with our model. First, to be consistent, 
we need to adopt the cooling function due to Raymond (Raymond~et~al.\ 
1976; Raymond \& Smith 1977) that was used by RTV78:%
\footnote
{Note that, to derive their scaling, RTV78 actually used an 
approximation of the Raymond cooling function: $Q(T)\sim T^{-1/2}$.}
\beq
Q^{\rm RTV}(T) \simeq 
2\cdot 10^{-22}\, T_6^{-2/3}\, {\rm erg}\, {\rm cm}^3\, {\rm s}^{-1} \, ,
\label{eq-Q-RTV}
\eeq
applicable for 2--10~MK (see Appendix~A of~RTV78).
Substituting this function into scalings~(\ref{eq-n*})--(\ref{eq-H*}), 
we get 
\begin{eqnarray}
n_* &\simeq& 
10^{10}\, {\rm cm}^{-3}\ B_{z,2}^{4/9}\, L_{9}^{-1/9}\, 
T_{\rm loop,6}^{10/9}\ (\gamma\tau_{\rm rad,0})^{2/3}      \, , 
\label{eq-n*-RTV-Q}   \\
B_{0,*} &\simeq& 
40\,{\rm G}\ B_{z,2}^{5/9}\, L_9^{1/9}\, 
T_{\rm loop,6}^{5/9}\ (\gamma\tau_{\rm rad, 0})^{1/3} \, ,
\label{eq-B_0*-RTV-Q}   \\
\Delta t_* &\simeq& 
160\ {\rm sec}\ B_{z,2}^{-4/9}\,L_{9}^{1/9}\,
T_{\rm loop,6}^{5/9}\ (\gamma\tau_{\rm rad,0})^{-2/3} \, ,
\label{eq-delta-t*-RTV-Q}   \\
E_* &\simeq& 
5\cdot 10^{28}\, {\rm erg}\ L_{\parallel,9}\, L_9^{2+2/9}\, B_{z,2}^{10/9}\, 
T_{\rm loop,6}^{10/9}\ (\gamma\tau_{\rm rad,0})^{2/3}\, ,
\label{eq-E*-RTV-Q}   \\
H_* &\simeq& 
3\cdot 10^{26}\, {\rm erg}\ {\rm sec}^{-1}\, L_{\parallel,9}\, L_9^{2+1/9}\, 
B_{z,2}^{14/9} \, \times \nonumber   \\ 
&\times&\, T_{\rm loop,6}^{5/9}\ (\gamma\tau_{\rm rad,0})^{4/3}\,.
\label{eq-H*-RTV-Q}
\end{eqnarray}

Next, recasting the RTV loop-temperature scaling in terms of 
the electron density instead of the pressure,
\beq
T_{\rm loop}^{\rm RTV} \simeq 
3\cdot 10^6\ {\rm K}\ n_{10}^{1/2}\, L_{\parallel,9}^{1/2} \, ,
\label{eq-RTV}
\eeq
and substituting it into the above scalings, we get:
\begin{eqnarray}
n_*^{\rm RTV} &\simeq& 
6\cdot 10^{10}\, {\rm cm}^{-3}\ B_{z,2}\, L_{9}^{-1/4}\, 
L_{\parallel,9}^{5/4}\, \gamma_{-3}^{3/2}      \, , 
\label{eq-n*-RTV}   \\
T_{\rm loop,*}^{\rm RTV} &\simeq& 
7\cdot 10^6\, {\rm K}\ B_{z,2}^{1/2}\, L_{9}^{-1/8}\, 
L_{\parallel,9}^{9/8}\ \gamma_{-3}^{3/4}      \, , 
\label{eq-T*-RTV}   \\
B_{0,*}^{\rm RTV} &\simeq& 
80\,{\rm G}\ B_{z,2}^{5/6}\, L_9^{1/24}\, 
L_{\parallel,9}^{5/8}\, \gamma_{-3}^{3/4} \, ,
\label{eq-B_0*-RTV}   \\
\Delta t_*^{\rm RTV} &\simeq& 
900\ {\rm sec}\ B_{z,2}^{-1/6}\, L_{9}^{1/24}\,
L_{\parallel,9}^{5/8}\, \gamma_{-3}^{-1/4} \, ,
\label{eq-delta-t*-RTV}   \\
E_*^{\rm RTV} &\simeq& 
2.5\cdot 10^{29}\, {\rm erg}\ L_{\parallel,9}^{9/4}\, L_9^{2+1/12}\, 
B_{z,2}^{5/3}\, \gamma_{-3}^{3/2}\, ,
\label{eq-E*-RTV}   \\
H_*^{\rm RTV} &\simeq& 
3\cdot 10^{26}\, {\rm erg}\ {\rm sec}^{-1}\, \times \nonumber \\
&\times&\, L_{\parallel,9}^{13/8}\, L_9^{2+1/24}\, B_{z,2}^{11/6}\, 
\gamma_{-3}^{7/4}\, ,
\label{eq-H*-RTV}
\end{eqnarray}
where, in addition, we introduced $\gamma_{-3}\equiv 
\gamma\cdot 10^3\,{\rm sec}$ and used $\tau_{\rm rad,0}=400$~sec.

A similar exercise could be conducted, for example, for an entirely 
different cooling function (Cook~et~al. 1989; Landi \& Landini 1999; 
Aschwanden~et~al. 2000a) and/or for a different scaling law [e.g., 
that derived by Cargill~et~al. (1995) or, observationally, by Kano 
\& Tsuneta (1995)].

These estimates give us the characteristic scales of micro- and nanoflare-like 
energy-release events in a bright coronal loop of a fixed length~$L_\parallel$,
fixed cross-loop spatial scale~$L$, and a fixed axial magnetic field~$B_z$, 
subject to continuous braiding at a fixed rate~$\gamma$. It is of course 
understood that this is just an idealized model set-up and that, in practice, 
using such typical values straight-forwardly may not be meaningful
because of the highly inhomogeneous nature of the solar corona.


\section{Discussion: What Needs to be Done}
\label{sec-discussion}

In this section I will discuss several open questions relevant to 
the physical picture presented in this paper --- questions that I 
would like to see answered in the near future. Correspondingly, I 
will describe possible studies (mostly numerical) that I think ought 
to be performed in order to confirm, modify, or refute various elements 
of my model.

First, the whole picture hinges, in part, on the premise that reconnection 
is slow in the collisional regime. Although in the above discussion I have 
for definiteness assumed that it is as slow as in the classical Sweet--Parker 
model, I don't actually think that this needs to be the case. If the true 
rate of classical resistive-MHD reconnection turns out to be much faster 
than Sweet--Parker, the overall picture would still stand, qualitatively, 
as long this rate is still much smaller than the collisionless reconnection 
rate. The only thing that matters is that there are two regimes, slow and 
fast, and the transition between them is governed by the collisionality of 
the reconnection layer. 

Nevertheless, from the point of view of my picture, studies of slow 
collisional (resistive MHD) reconnection are just as important as those
of fast collisionless reconnection. One would like to tighten up the case 
for slow reconnection. Thus, I would like to encourage more studies of 
resistive reconnection, to really confirm that it is slow. Thus, it is 
important to address the following specific issues, best using numerical 
simulations:

1) Does the presence of small-scale 3D MHD turbulence enhance reconnection
rate, as suggested by Lazarian \& Vishniac (1999)?  If it does, how large
is the enhancement and what determines it? Could it be as fast as 
collisionless reconnection, in which case it would constitute a challenge 
to the picture presented in this paper? To address this issue, one would 
have to perform a 3D MHD simulation involving a regular large-scale 
reconnection layer with a super-imposed 3D MHD turbulence. Just as 
importantly, one needs to investigate the possibility that 3D MHD turbulence 
is spontaneously generated inside the layer by the secondary-instability 
mechanism of Dahlburg~et~al., (1992, 2005), and evaluate its effect on 
the time-averaged reconnection rate.

2) What is the effect of the 2D tearing instability inside the resistive 
Sweet--Parker layer, on the overall, time-averaged reconnection rate? 
It is expected that tearing instability may become important in 
reconnection layers that are very long, with aspect ratios $L/\delta>100$.
It may then make the reconnection process inherently time-dependent, bursty. 
Therefore, in order to investigate this issue, one would have to perform 
only a 2D simulation, but with a very high resolution (corresponding 
to~$S> 10^4$) and with a very long duration (to be able to average over 
many bursts).

3) The majority of existing numerical studies of resistive-MHD reconnection 
use a constant uniform resistivity. It is important to check whether the main 
results will be the same if one uses the actual Spitzer resistivity.
This is especially important in the context of a magnetically-dominated
environment, such as the solar corona, where the temperature inside the
reconnection layer may be much higher than the ambient temperature. 
As a result, the Spitzer resistivity may vary by a factor of a hundred 
across the layer, being much smaller at its center.

4) One should further pursue numerical studies of magnetic reconnection 
that include both resistive MHD and Hall effect, along the lines of the
recent work by Cassak~et~al.\ (2005, 2006, 2007). Of particular interest 
is the intermediate regime, when $\delta_{\rm SP} > d_i,\rho_i$ but at 
the same time $\Omega_e >\nu_e$. The objective of such simulations would 
be to observe the transition from slow to fast reconnection within one
simulation run as the upstream plasma parameters, such as~$B_0$ or~$n$,
are gradually changed, passing the critical point. Whereas the results 
published by Cassak~et~al.\ (2005, 2006, 2007) are already very important, 
the relevant parameter space is large and similar studies need to be extended 
to other regimes, most notably to a situation where $\beta_{\rm upstream} 
\ll 1$. Also, these results need to be confirmed by other groups, preferably 
with a more realistic electron-to-ion mass ratio, a larger box size, and  
higher numerical resolution.

5) Future numerical studies of collisional reconnection should include
realistic energy balance taking into account both Ohmic heating and the
electron thermal conduction. The goal here would be to calculate the
values of~$T_e$ and~$n_e$ at the center of the reconnection region.
Also, one needs to study temperature equilibration between ions and 
electrons.

6) Effect of viscosity on resistive-MHD reconnection needs to be assessed.

7) The effect of a strong guide field on resistive-MHD reconnection
and on the transition to the fast regime (e.g., Cassak et al.\ 2007)
needs to be investigated in more detail.

8) Finally, I would like to strongly encourage further  experimental 
(laboratory) studies of collisional reconnection, especially in the 
large-$S$ limit.

In addition to the above questions related to collisional reconnection,
there are several important issues related to collisionless reconnection:
(1) What is its physical nature of anomalous resistivity due to wave-particle 
interactions and under what conditions is it excited? 
(2) What is the structure of the Petschek-like reconnection layer 
for a given functional form of anomalous resistivity and what is 
the resulting reconnection rate in terms of the basic physical parameters? 
Does the anomalous resistivity spread along the separatrices (Petschek's 
shocks) or is it present only in the small central region? If the latter 
is the case, then how does the plasma crossing the shocks gets heated? 
(3) When does laminar a Hall-effect (or, in general, two-fluid) 
reconnection take place? How do Hall effect and anomalous resistivity 
co-exist and interact with each other?
(4) How rapid is two-fluid reconnection? What parameters affect 
the reconnection rate in this regime? 
(5) What is the effect of guide field on triggering and saturation
of Hall-regime reconnection and on anomalous resistivity?
(6) Is collisionless reconnection process bursty and, if so, what is 
the time-averaged reconnection rate?
(7) What is the overall partitioning of the released energy between 
the bulk kinetic energy, electron and ion thermal energies, and 
nonthermal particle acceleration?

Finally, in order to be realistic, future numerical simulations 
of the solar corona should include all of the following (c.f., 
Miyagoshi \& Yokoyama 2003; Klimchuk~2006): 
(1) flux emergence processes and random motions of the field-line footpoints;
(2) physically-motivated prescription for transition to fast reconnection,
such as the one suggested in this paper; such a prescription would thus
play a role of a subgrid model used in a large-scale MHD simulation of 
the corona;
(3) mass exchange between the corona and the solar surface
(e.g., chromospheric evaporation and plasma precipitation); 
(4) optically-thin radiative energy losses and thermal conduction 
(including the contribution due to nonthermal electrons) along 
the magnetic field lines.


\section{Conclusions}
\label{sec-conclusions}

Magnetic reconnection research started 50 years ago in the field of 
Solar Physics, with the Sweet--Parker (Sweet~1958; Parker~1957) model
for solar flares, followed by the Petschek (1964) theory a few years 
later. These studies tackled the most fundamental micro-physical aspects 
of the reconnection layer. Over time, however, the focus of solar 
reconnection research has shifted away from local basic physics of 
reconnection, despite the fact that reconnection has been confirmed 
observationally to be the key process responsible for magnetic energy 
release in flares (e.g., Tsuneta~1996, Yokoyama~et~al. 2001). 
Instead, the forefront of research on the fundamental physics of 
reconnection has moved in recent years to other fields, most notably, 
to Space Physics, where detailed in-situ measurements using Earth-orbiting 
spacecraft are now available (e.g., Oieroset~et~al. 2001; Nagai~et~al. 2001; 
Mozer~et~al. 2002), and to Laboratory Plasma Physics, where several dedicated 
experiments (Yamada~et~al.\ 1997; Egedal~et~al.\ 2007; Brown~1999) have 
already made fundamental contributions to our understanding of reconnection. 
Most importantly, a lot of progress has been recently made using numerical 
simulations, again, mostly in the context of the Earth Magnetosphere.

The basic paradigm that emerges as a result of all these studies can 
be summarized as follows (see~\S~\ref{subsec-2regimes}). The starting
point is the realization that there are indeed two reconnection regimes. 
The first one is a slow (Sweet-Parker) resistive-MHD regime that is 
realized in relatively dense, collisional plasmas. The second one is 
a fast (Petschek-like) regime that takes place in collisionless plasmas. 
The mechanism for the fast collisionless reconnection can be either a 
locally-enhanced anomalous resistivity due to micro-instabilities triggered 
when the current density exceeds a certain threshold; or the Hall effect. 
In either case, one can formulate an approximate condition for the transition 
between the slow collisional and the fast collisionless regimes. If the guide 
field is not much larger than the reconnecting magnetic field, this condition 
is $\delta_{\rm SP} < d_i$, where $\delta_{\rm SP}$ is the thickness of the 
Sweet--Parker reconnection layer and $d_i$ is the collisionless ion skin 
depth. One can further rewrite this condition in terms of the classical 
electron mean free path~$\lambda_{e,\rm mfp}$ inside the layer as $L<L_c
\sim (m_i/m_e)^{1/2}\, \lambda_{e,\rm mfp}$, where~$L$ is the global system 
size (Yamada~et~al.~2006). Due to the strong temperature dependence of 
$\lambda_{e,\rm mfp}$, this form of the condition highlights the need to 
estimate the electron temperature~$T_e$ at the center of the layer. 
Using considerations of pressure balance and energy conservation
(see~\S~\ref{subsec-T_e-n_e-SP}), one can express~$T_e$ in terms 
of the reconnecting magnetic field~$B_0$ and the background plasma 
density~$n$. Then, the collisionless reconnection condition can be 
recast in terms of~$L$, $B_0$ and~$n$, see equation~(\ref{eq-L_c-final}) 
in~\S~\ref{subsec-condition}. In the case with a strong guide field~$B_z
\gg B_0$, the corresponding condition is $\delta_{\rm SP}<\rho_i$ and 
this leads to an additional factor~$(B_0/B_z)^2$ in the expression for 
the critical length~$L_c$ [see eq.~(\ref{eq-L_c-guide})].

One of the main driving forces behind this paper is the author's desire 
to bring the recently-obtained knowledge about reconnection back to Solar
Physics and to use it productively to build a better understanding of the 
solar corona. Although most of the present discussion is also relevant 
to solar flares, in this paper I focus predominantly on the problem of 
solar coronal heating (see~\S~\ref{sec-corona}).

In the context of Parker's (1983, 1988) nanoflare theory of coronal heating,
magnetic energy release in the solar corona takes place in the form of many 
unresolved reconnection events (nanoflares). One of the most important 
features of this picture is the intermittent character of energy release. 
Random footpoint motions lead to continuous twisting of elementary magnetic 
strands around each other, which, in turn, leads to the formation of many 
small current sheets in the corona. Current sheets may form either in finite 
time, as suggested by Parker (1983, 1986), or exponentially in time, as was 
demonstrated by van~Ballegooijen (1986) and later numerically by Galsgaard 
\& Nordlund (1996); for our purposes, it does not matter which one is correct.
What matters is that thin current layers do eventually form. Free magnetic 
energy accumulates for a while and is then suddenly released in distinct 
fast reconnection events.
In the present paper, I suggest that the transition between the slow and 
fast reconnection regimes plays a key role in determining when a given 
nanoflare will take off and how much energy will be released (see also 
Cassak~et~al. 2005). My model can thus be regarded as an alternative to 
the secondary-instability mechanism proposed by Dahlburg~et~al.\ (2003).
Furthermore, I argue that the fact that the fast reconnection condition 
involves the ambient plasma density is an important part of the story. 
The reason for this is that the density in the corona is not a fixed 
constant; in a given loop, it constantly changes in response to radiative 
cooling and chromospheric evaporation caused by coronal energy-release events. 
The basic picture here is the following. A coronal energy release leads to an 
increase in density, thus making the plasma more collisional. This temporarily 
inhibits fast reconnection in the given region until the density decreases 
again (on the radiative cooling timescale) to below a certain critical value. 
At this point fast reconnection again becomes possible. On the longer 
time-scale, a quasi-periodic behavior is established, characterized 
by repeated cycles that include fast reconnection events, followed by 
chromospheric evaporation episodes, followed by relatively long 
($\sim$ 1~hour) and steady periods during which free magnetic energy 
in the loop builds up and the plasma gradually cools down. Thus, coronal 
heating can be viewed as a self-regulating process that statistically keeps 
the density roughly near the critical value for the fast reconnection 
transition. In other words, the system constantly fluctuates around 
the state of marginal collisionality as defined by the collisionless 
reconnection condition. In the long-term statistical equilibrium, 
a balance is maintained in which the amount of plasma pumped into 
a coronal loop as a result of an evaporation episode is equal to 
the amount of plasma drained down to the surface during the gradual 
radiative cooling stage that takes place between two subsequent 
fast-reconnection events. The characteristic equilibrium density, 
the time interval between reconnection episodes, and the energy 
released in each such episode can be estimated in terms of the loop's 
longitudinal magnetic field, its characteristic size, and the footpoint 
driving rate (see~\S~\ref{sec-corona}).

Finally, I believe that the physical framework developed in this paper 
should also be applicable to magnetically-dominated coronae of other 
astrophysical objects, such as other stars and accretion disks 
(Goodman \& Uzdensky 2007).



\begin{acknowledgments}

I am grateful to S.~Antiochos, P.~Cassak, J.~Goodman, H.~Ji, J.~Klimchuk, 
R.~Kulsrud, E.~Parker, M.~Shay, and M.~Yamada for stimulating discussions 
and encouraging remarks and to the anonymous referee for useful suggestions.

This work is supported by National Science Foundation Grant 
No.\, PHY-0215581 (PFC: Center for Magnetic Self-Organization 
in Laboratory and Astrophysical Plasmas).

\end{acknowledgments}


\section*{REFERENCES}
\parindent 0 pt


Aschwanden, M.~J, et al.\ 2000a, ApJ, 535, 1047

Aschwanden, M.~J., Nightingale, R.~W., \& Alexander, D.\ 2000b, ApJ, 541, 1059

Aschwanden, M.~J., Schrijver, C.~J., \& Alexander, D.\ 2001a, ApJ, 550, 1036

Aschwanden, M.~J., Poland, A.~I., \& Rabin, D.~M.\ 2001b, ARA\&A, 39, 175

Aschwanden, M.~J., Winebarger, A., Tsiklauri, D., Peter, H. 2007, 
ApJ, 659, 1673

Bhattacharjee, A., Ma, Z.~W., \& Wang, X.\ 1999, JGR, 104, 14543

Bhattacharjee, A., Ma, Z.\ W., \& Wang, X.\ 2001, Phys.\ Plasmas, 8, 1829

Birn, J.\ et al.\ 2001, \jgr, 106, 3715

Biskamp, D.\ 1986, Phys. Fluids, 29, 1520.

Biskamp, D., Schwarz, E., \& Drake, J.~F.\ 1995, PRL 73, 3850

Biskamp, D., Schwarz, E., \& Drake, J.~F.\ 1997, Phys.\ Plasmas, 4, 1002 

Biskamp, D., \& Schwarz, E.\ 2001, Phys.\ Plasmas, 8, 4729

Breslau, J.~A. \& Jardin, S.~C.\ 2003, Phys. Plasmas, 10, 1291

Brown, J.~C., Krucker, S., G\"udel, M., \& Benz. A.~O. 2000, A\&A, 359, 1185

Brown, M.~R. 1999, Phys. Plasmas, 6, 1717

Brown, M.~R., Cothran, C.~D., \& Fung, J. 2006, Phys. Plasmas, 13, 056503

Bulanov, S.~V., Syrovatsky, S.~I., \& Sakai, J.\ 1978, JETP Lett., 28, 177

Cargill, P.~J., Mariska, J.~T., \& Antiochos, S.~K.\ 1995, ApJ, 439, 1034

Cassak, P., Shay, M., \& Drake, J. 2005, \prl, 95, 235002

Cassak, P., Drake, J., \& Shay, M. 2006, ApJ, 644, L145

Cassak, P., Drake, J., \& Shay, M. 2007, Phys. Plasmas, 14, 054502

Cook, J.~W., Cheng, C.-C., Jacobs, V.~L., \& Antiochos, S.~K. 1989,
ApJ, 338, 1176

Coppi, B. \& Frieldland, A.~B. 1971, ApJ, 169, 379

Coroniti, F.~V. \& Eviatar, A. 1977, ApJS, 33, 189

Dahlburg, R.~B., Antiochos, S.~K., \& Zang, T.~A. 1992, Phys. Fluids B, 4, 3902

Dahlburg, R.~B., Klimchuk, J.~A., \& Antiochos, S.~K. 2005, ApJ, 622, 1191

Daughton, W., Scudder, J., \& Karimabadi, H.\ 2006, Phys. Plasmas, 13, 072101

Egedal, J., Fox, W., Katz, N., Porkolab, M., Reim, K., \& Zhang, E. 2007,
Phys. Rev. Lett., 98, 015003

Erkaev, N.~V., Semenov, V.~S., \& Jamitzky, F.\ 2000, 
Phys. Rev. Lett., 84, 1455 
	
Erkaev, N.~V., Semenov, V.~S., Alexeev, I.~V., \& Biernat, H.~K.\ 2001,
Phys. Plasmas, 8, 4800

Forbes, T.~G. \&  Priest, E.~R.\ 1983, Solar Phys., 84, 169

Fujimoto, K.\ 2006, Phys.\ Plasmas, 13, 072904

Galsgaard, K. \& Nordlund, A.\ 1996, JGR, 101, 13445

Goodman, J. \& Uzdensky, D.~A. 2007, in preparation

Gudiksen, B.~V. \& Nordlund, A.\ 2002, ApJ, 572, L113

Gudiksen, B.~V. \& Nordlund, A.\ 2005, ApJ, 618, 1020

Heitsch, F. \& Zweibel, E.~G.\ 2003, ApJ, 583, 229

Hendrix, D.~L., van Hoven, G., Mikic, Z., \& Schnack, D.~D.\ 
1996, ApJ, 470, 1192 

Hsu S.~C., Carter, T.~A., Fiksel, G., Ji, H., Kulsrud, R.~M., \& Yamada, M.\
2001, Phys. Plasmas, 8, 1916

Huba, J.~D. \& Rudakov, L.~I.\ 2004, Phys. Rev. Lett., 93, 175003 

Hesse, M., Schindler, K., Birn, J., \& Kuznetsova, M.\ 1999, 
Phys. Plasmas, 6, 1781

Ji, H., Yamada, M., Hsu S., \& Kulsrud, R. 1998, Phys. Rev. Lett., 80, 3256
	
Ji, H., Terry, S., Yamada, M., Kulsrud, R. Kuritsyn, A., \& Ren, Y.\
2004, Phys. Rev. Lett., 92, 115001 

Kano, R. \& Tsuneta, S. 1995, ApJ, 454, 934 

Karimabadi, H., Daughton, W., \& Scudder, J.\ 2007, 
Geophys. Res. Lett., 34, L13104
	
Kim, E.-J., \& Diamond, P.~H. 2001, Phys. Lett. A, 291, 407 

Kleva, R.~G., Drake, J.~F., \& Waelbroeck F.~L., 1995,
Phys. Plasmas, 2, 23

Klimchuk, J.~A.\ 2006, Solar Phys., 234, 41

Kulsrud, R.~M.\ 2001, Earth, Planets and Space, 53, 417

Kulsrud, R.~M.\ 2005, ``Plasma Physics for Astrophysics'',
Princeton Univ. Press, Princeton 

Kulsrud, R., Ji, H., Fox, W., \& Yamada, M. 2005, Phys. Plasmas, 12, 082301

Kuritsyn, A., Yamada, M., Gerhardt, S., Ji, H., Kulsrud, R., \& Ren, Y., 
2006, Phys. Plasmas, 13, 055703

Landi, E. \& Landini, M. 1999, A\&A, 347, 401

Lee, L.~C. \&  Fu, Z.~F. 1986, J. Geophys. Res., 91, 6807

Lottermoser, R.-F. \& Scholer, M.\ 1997, JGR, 102, 4875

Loureiro, N.~F., Schekochihin, A.~A., \& Cowley, S.~C.\ 2007, 
submitted to PRL; preprint (astro-ph/0703631)

Lyutikov, M. \& Uzdensky, D. 2003, ApJ, 589, 893

Ma, Z.~W. \& Bhattacharjee, A. 1996, Geophys. Res. Lett., 23, 1673 

Malyshkin, L.~M., Linde, T., \& Kulsrud, R.~M.\ 2005, Phys. Plasmas, 12, 102902

Masuda, S., Kosugi, T., Hara, H., Tsuneta, S., \& Ogawara, Y.\ 1994,
Nature, 371, 495

Masuda, S., Kosugi, T., Hara, H., Sakao, T., Shibata, K., \& Tsuneta, S.\ 
1995, PASJ, 47, 677

Miyagoshi, T. \& Yokoyama, T. 2003, ApJ, 593, L133

Mozer F.~S., Bale, S.~D., \& Phan, T.~D.\ 2002, 
Phys.\ Rev.\ Lett., 89, 015002

Nagai, T., Shinohara, I., Fujimoto, M., Hoshino, M., Saito, Y., 
Machida, S., \& Mukai, T.\ 2001, JGR, 106, 25929 

Oieroset, M., Phan, T.D., Fujimoto, M., Lin, R.P., \& Lepping, R.P.\
2001, Nature, 412, 414

Parker, E.~N.\ 1957, \jgr, 62, 509

Parker, E.~N.\ 1972,  ApJ, 174, 499

Parker, E.~N.\ 1983, \apj, 264, 642

Parker, E.~N.\ 1988, \apj, 330, 474

Petschek, H.~E.\ 1964, AAS-NASA Symposium on Solar Flares, 
(National Aeronautics and Space Administration, Washington, 
DC, 1964), NASA SP50, 425.

Porter, L. \& Klimchuk, J.\ 1995, ApJ, 454, 499

Priest, E.\ 1984, Solar Magnetohydrodynamics, (Reidel, Dordrecht, 1984)

Raymond, J.~C., Cox, D.~P., \& Smith, B.~W. 1976, ApJ, 204, 290 

Raymond, J.~C. \& Smith, B.~W. 1977, ApJS, 35, 419 

Ren, Y., Yamada, M., Gerhardt, S., Ji, H., Kulsrud, R., \& Kuritsyn, A.,
2005, Phys. Rev. Lett., 95, 055003
 
Rogers, B.~N., Denton, R.~E., Drake J.~F., \& Shay, M.~A.\ 2001, 
Phys. Rev. Lett. 87, 195004

Rosner, R., Tucker, W.~H., \& Vaiana, G.~S.\ 1978, ApJ, 220, 643 (RTV78)

Sato, T.\ \& Hayashi, T.\ 1979, Phys. Fluids, 22, 1189

Scholer, M.\ 1989, \jgr, 94, 8805.

Shay, M.~A. \& Drake, J.~F.\ 1998, JGR, 103, 9165

Shay, M.~A., Drake, J.~F., Rogers, B.~N., \& Denton, R.~E.\ 1999,
Geophys. Res. Lett., 26, 2163

Shay, M.~A., Drake, J.~F., Rogers, B.~N., \& Denton, R.~E.\ 2001,
JGR, 106, 3759

Shay, M.~A., Drake, J.~F., \& Swisdak, M. 2007, submitted to PRL; 
preprint (arxiv:0704.0818)

Shibata, K., Masuda, S., Shimojo, M., Hara, H., Yokoyama, T., 
Tsuneta, S., Kosugi, T., Ogawara, Y.\ 1995, ApJ, 451, L83

Shibata, K. 1996, Adv. Space Res., 17(4/5), 9 

Smith, P.~F. \& Priest, E.~R. 1972, ApJ, 176, 487

Strauss, H.~R., 1988, ApJ, 326, 412

Sweet, P.~A.\ 1958, in IAU Symp.~6, Electromagnetic Phenomena in Cosmical 
Physics, ed.\ B.~Lehnert, (Cambridge: Cambridge Univ. Press), 123.

Thompson, C.\ 1994, MNRAS, 270, 480

Titov, V.~S. \& Hornig, G.\ 2002, Adv.~Space Res., 29, 1087

Trintchouk, F., Yamada, M., Ji, H., Kulsrud, R.~M. \& Carter, T.~A.\ 
2003, Phys. Plasmas 10, 319

Tsuneta, S., Takakura, T., Nitta, N., Makishima, K., Murakami, T., Oda, M.,
Ogawara, Y., Kondo, I., Ohki, K., \& Tanaka, K.\ 1984, ApJ, 280, 887

Tsuneta, S., Hara, H., Shimizu, T., Acton, L.~W., Strong, K.~T., 
Hudson, H.~S., \& Ogawara, Y.\ 1992, PASJ, 44, L63

Tsuneta, S.\ 1996, ApJ,  456, 840 

Ugai, M., \& Tsuda, T.\ 1977, J. Plasma Phys., 17, 337.

Ugai, M.\ 1986, Phys. Fluids, 29, 3659

Ugai, M.\ 1992, Phys. Fluids B, 4, 2953

Ugai, M.\ 1999, Phys. Plasmas, 6, 1522

Uzdensky, D.~A., \& Kulsrud, R.~M.\ 1998, Phys. Plasmas, 5, 3249

Uzdensky, D.~A., \& Kulsrud, R.~M.\ 2000, Phys. Plasmas, 7, 4018.

Uzdensky, D.~A.\ 2003, ApJ, 587, 450

Uzdensky, D.~A. \& Kulsrud, R.~M.\ 2006, Phys. Plasmas, 13, 062305

Uzdensky, D.~A.\ 2006, ArXiv Astrophysics e-prints, astro-ph/0607656

Uzdensky, D.~A.\ 2007, Mem.~Soc.~Ast.~It., 78, 317; 
ArXiv Astrophysics e-prints: astro-ph/0702699 

Uzdensky, D.~A. \& MacFadyen, A.~I.\ 2006, ApJ, 647, 1192

van Ballegooijen, A.~A.\ 1986, ApJ, 311, 1001


Yamada, M., Ji, H., Hsu, S., Carter, T., Kulsrud, R., Bretz, N., 
Jobes, F., Ono, Y., \& Perkins, F. 1997, Phys. Plasmas, 4, 1936

Yamada, M., et al.\  2006, Phys. Plasmas, 13, 052119

Yokoyama, T., Akita, K., Morimoto, T., Inoue, K., \& Newmark, J.\ 
ApJ, 546, L69


\end{document}